\documentclass[journal]{IEEEtran}

\usepackage{amsmath,amssymb,amsfonts,amsthm}
\usepackage{bm}
\usepackage{graphicx}
\usepackage{cite}
\usepackage[T1]{fontenc}
\usepackage{hyperref}
\hypersetup{hidelinks}
\usepackage{flafter}
\usepackage{placeins}
\newcommand{\balpha}{\boldsymbol{\alpha}}
\newcommand{\btau}{\boldsymbol{\tau}}
\newcommand{\bomega}{\boldsymbol{\omega}}

\DeclareMathOperator{\sinc}{sinc}
\DeclareMathOperator*{\rank}{rank}
\DeclareMathOperator*{\essinf}{ess\,inf}
\DeclareMathOperator*{\esssup}{ess\,sup}

\theoremstyle{plain}
\newtheorem{thm}{Theorem}[section]
\newtheorem{lem}{Lemma}[section]
\theoremstyle{remark}
\newtheorem{rem}{Remark}[section]
\theoremstyle{definition}
\newtheorem{example}{Example}[section]
\numberwithin{equation}{section}

\begin{document}

\title{Multichannel Sampling and Reconstruction in Multidimensional Shift-Invariant Spaces}

\author{Riya Ghosh, Satish Mulleti, and Chandra Sekhar Seelamantula%
\thanks{Riya Ghosh and Satish Mulleti are with the Department of Electrical Engineering, Indian Institute of Technology Bombay, Mumbai, India (e-mail: riya74012@gmail.com; mulleti.satish@gmail.com).}%
\thanks{Chandra Sekhar Seelamantula is with the Department of Electrical Engineering, Indian Institute of Science, Bengaluru, India (e-mail: css@iisc.ac.in).}}

\markboth{}%
{Ghosh \MakeLowercase{\textit{et al.}}: Multichannel Sampling and Reconstruction in Multidimensional Shift-Invariant Spaces}

\maketitle

\begin{abstract}
We develop a multichannel sampling and reconstruction framework for multidimensional principal shift-invariant spaces based on a finite-dimensional polyphase representation. Periodic nonuniform shifts and derivative measurements are modeled as analysis channels, leading to a frequency-domain linear system relating the measurements to the polyphase components of the underlying coefficient sequence. Stable recovery is characterized through uniform singular-value bounds on the associated polyphase matrix, and explicit synthesis kernels are obtained from its Moore–Penrose left inverse. The formulation applies to both separable and genuinely non-separable generators and includes several classical one-dimensional sampling formulas as special cases. For irregular derivative measurements, where the periodic polyphase structure is no longer directly available, we develop an iterative reconstruction method based on local bivariate Hermite interpolation followed by orthogonal projection onto the shift-invariant space. A sufficient contraction condition is derived to guarantee geometric convergence of the iteration. Numerical experiments demonstrate the reconstruction performance for separable and non-separable generators and illustrate robustness to additive measurement noise and sampling-location perturbations.
\end{abstract}

\begin{IEEEkeywords}
Derivative sampling, Hermite interpolation, iterative reconstruction, multichannel sampling, polyphase matrix, shift-invariant spaces.
\end{IEEEkeywords}

\section{Introduction}
Sampling theory provides the mathematical foundation for recovering continuous signals from discrete measurements and plays a central role in signal processing, harmonic analysis, and approximation theory. Beginning with the classical Shannon sampling theorem for bandlimited functions \cite{sha49,unser}, sampling theory has subsequently been extended to nonuniform sampling, generalized measurements, and broader signal models such as shift-invariant spaces.

Shift-invariant spaces provide a flexible alternative to classical bandlimited models and arise naturally in spline approximation, wavelet analysis, and signal representation. Aldroubi and Gröchenig \cite{Non-uniform_Siam,aldroubibeurling} developed fundamental results on nonuniform sampling and reconstruction in shift-invariant spaces. Their framework applies in arbitrary dimensions and to broad classes of generators, and also provides iterative reconstruction methods for sufficiently dense nonuniform sampling sets. Generalized, vector-valued, and multigenerator sampling in shift-invariant spaces was subsequently studied in several works; see, for example, \cite{Garcia,multigenerator2,vector_sampling}.

Sampling with derivatives forms another important extension of classical sampling theory. Early results include stable nonuniform sampling expansions involving derivatives and iterative reconstruction from derivative measurements \cite{Rawn,Razafinjatovo}. More recent work has addressed density conditions for sampling with derivatives or bunched measurements \cite{Derivative_1}, as well as derivative sampling in shift-invariant spaces, periodic shift-invariant spaces, and finitely generated shift-invariant spaces \cite{RiyaAntonyRiM1,Sharp_result_Grochenig,Antony_Riya_NFAA,Priyanka_selvan_IEEE,Derivative_approximation,Priyanka_Selvan_JCAM}. These results provide an important theoretical basis for reconstructing signals when function values and derivative measurements are jointly available.

Multivariate sampling in shift-invariant spaces has also been investigated in several settings. Sun \cite{Sun2005} studied regular and irregular sampling in multivariate shift-invariant subspaces. García and Pérez-Villalón \cite{GarciaPerezVillalon2009} developed a multivariate generalized sampling theory in which measurements obtained from convolution operators are acquired on suitable sublattices of the integer lattice. García, Muñoz-Bouzo, and Pérez-Villalón \cite{GarciaMunozPerez2011} subsequently derived regular multivariate generalized sampling formulas in $L^p$ shift-invariant spaces, including function values, derivatives, and filtered measurements. Related iterative reconstruction methods for multivariate irregular sampling in spline-like spaces were developed by Aldroubi and Feichtinger \cite{AldroubiFeichtinger1998}.

Multichannel and periodic nonuniform sampling constitute a closely related line of research. Papoulis' generalized sampling expansion \cite{Papouli} showed that a bandlimited signal can be reconstructed from several filtered measurement sequences. Unser and Zerubia \cite{Unser_Zerubia} later extended this viewpoint beyond the bandlimited setting and formulated generalized sampling in shift-invariant reconstruction spaces. In their $m$-channel formulation, the outputs of $m$ linear shift-invariant systems are sampled at one $m$-th of the reconstruction rate, illustrating the principal advantage of distributing the available information across several acquisition channels. Periodic nonuniform sampling in shift-invariant spaces was studied in detail by Hogan and Lakey \cite{Hogan2006}, who related this problem to the generalized-sampling viewpoint and analyzed stable reconstruction using tools such as the Zak transform. Multichannel sampling in one-dimensional shift-invariant spaces was further investigated, for example, by Kang et al. \cite{KangKimKwon2010} and García et al. \cite{GarciaKimKwonYoon2012}.

Building on the multivariate generalized sampling frameworks in \cite{GarciaPerezVillalon2009, GarciaMunozPerez2011}, we develop an explicit coefficient-domain polyphase formulation for periodic nonuniform mixed-derivative sampling in multidimensional principal shift-invariant spaces. The formulation makes the dependence of the sampling matrix on the generator, derivative orders, and sampling shifts explicit. It provides a common construction of sufficient stability conditions and synthesis kernels, which we illustrate through classical sampling formulas and multidimensional examples with separable and non-separable generators.

A further distinction concerns the structure of the generating function. Tensor-product constructions form an important class of multidimensional sampling models; see, for example, \cite{garcia2016modeling}. Such constructions inherit a coordinate-wise structure from the corresponding one-dimensional spaces. Our formulation does not require a tensor-product factorization of the generator and therefore applies equally to separable and genuinely non-separable generators. This distinction is illustrated numerically using both classical separable generators and strictly non-separable examples, including a sinc-Gaussian generator and a radial generator constructed from the Gauss hypergeometric function.

The second part of this paper concerns derivative sampling on irregular Cartesian grids in two dimensions. Complementing the one-dimensional interpolation–projection approach in \cite{Priyanka_Selvan_JCAM}, we derive a bivariate Hermite approximation bound that accounts separately for the errors in each coordinate and the contribution of mixed derivatives. Combining this estimate with a Bernstein-type inequality yields a sufficient convergence condition that explicitly relates the sampling gaps and interpolation orders to spectral properties of the generator. The mixed-derivative contribution accounts for coupling between the coordinates without requiring a tensor-product factorization of the generator. Under this condition, Hermite interpolation followed by orthogonal projection produces an iterative reconstruction with a quantitative geometric error bound.

The main contributions are as follows: (i) a multidimensional multichannel/polyphase model that incorporates periodic nonuniform, derivative, and filtered measurements; (ii) sufficient conditions for unique and stable recovery and explicit synthesis kernels obtained from the Moore--Penrose left inverse of the polyphase matrix; (iii) separable and genuinely non-separable examples showing how classical one-dimensional formulas arise as special cases; and (iv) for irregular derivative samples, an iterative reconstruction method based on local bivariate Hermite interpolation followed by orthogonal projection, together with a contraction condition, error estimates, and robustness experiments under measurement noise and sampling jitter.

Section II introduces the sampling model, Section III develops the polyphase reconstruction theory and examples, Section IV treats irregular derivative sampling and numerical experiments, and Section V concludes the paper.

\section{Preliminaries and Sampling Model}\label{sec: 2}
In this section, we introduce the foundational spaces, sampling model, and mathematical notation utilized throughout the paper.

\subsection{Notation and basic spaces}
Throughout the paper, boldface lowercase letters denote vectors in $\mathbb R^d$ or $\mathbb Z^d$. We denote by $\mathbb N_0=\mathbb N\cup\{0\}$ the set of nonnegative integers.
The space $L^2(\mathbb R^d)$ consists of all square-integrable functions on $\mathbb R^d$, equipped with the norm
$$ \|f\| = \left( \int_{\mathbb R^d}|f(\mathbf t)|^2\,d\mathbf t \right)^{1/2}. $$
For $f\in L^1(\mathbb R^d)\cap L^2(\mathbb R^d)$, we use the Fourier-transform convention
$$ \widehat f(\boldsymbol\omega) = \int_{\mathbb R^d} f(\mathbf t)e^{-2\pi \mathrm{j}\boldsymbol\omega\cdot\mathbf t}\,d\mathbf t, $$ where $\bomega \cdot \mathbf{t}$ is the Euclidean dot product between $\bomega$ and $\bf t$. 
The sequence space $\ell^2(\mathbb Z^d)$ consists of all square-summable sequences
$\mathrm{c}=\{c[\mathbf k]\}_{\mathbf k\in\mathbb Z^d}$
on $\mathbb Z^d$, with
$$ \|\mathrm c\|_{\ell^2} = \left( \sum_{\mathbf k\in\mathbb Z^d}|c[\mathbf k]|^2 \right)^{1/2} <\infty. $$
For vectors $\mathbf a=(a_1,\ldots,a_d)$ and $\mathbf b=(b_1,\ldots,b_d)$, we denote their componentwise product by
\begin{align}\label{odot}
     \mathbf a\odot\mathbf b := (a_1b_1,\ldots,a_db_d). 
\end{align}

For a multi-index
$\boldsymbol\alpha=(\alpha_1,\ldots,\alpha_d)\in\mathbb N_0^d,$
we use
\begin{align}\label{notation_2D}
     |\boldsymbol\alpha| = \alpha_1+\cdots+\alpha_d, \quad \mathcal D^{\boldsymbol\alpha} = \frac{\partial^{|\boldsymbol\alpha|}} {\partial t_1^{\alpha_1}\cdots\partial t_d^{\alpha_d}},
\end{align}
and
\begin{align}\label{notation_2D(1)}
    (2\pi \mathrm{j}\bm\omega)^{\boldsymbol{\alpha}}=\prod_{i=1}^d(2\pi \mathrm{j}\omega_i)^{\alpha_i}.
\end{align}
A $d$-dimensional shift operator $\mathcal{S}_{\btau}$ is defined as $\mathcal{S}_{\btau}f(\bf t)  := f(\mathbf{t} - \btau)$. 
Let $\mathbf{P}=(P_1, \cdots, P_d)$ where $P_i\in \mathbb{N}$ for $1\le i\le d$. Define an index set
$$\mathcal I:=\left\{
\bm{l}=(l_1,\ldots,l_d)\in\mathbb N_0^d:~
0\le l_i\le P_i-1,\;1\le i\le d
\right\}.$$ 
For each $\bm l\in\mathcal I$, define the corresponding polyphase component by
$$x_{\bm l}[\mathbf r]
=x[\mathbf P\odot\mathbf r+\bm l],\quad \mathbf r\in\mathbb Z^d.$$
Then the original sequence can be expressed in terms of its polyphase components as
\begin{align}
  x[\bm{n}] =  \sum_{\bm{l}\in \mathcal{I}}  \sum_{\mathbf{r}\in \mathbb{Z}^d}x_{\bm l}[\mathbf{r}]\, \delta[\bm{n}-(\mathbf{P}\odot \mathbf{r}+\bm l)], \label{eq:d-polyphase}
\end{align}
where $\delta[\bm{n}]$ denotes the $d$-dimensional Kronecker delta. There are $P = \prod_{i=1}^d P_i$ polyphase components.

\subsection{Shift-invariant spaces (SISs)}
We model our continuous signals using principal SISs. The space generated by a basis function $\phi \in L^2(\mathbb{R}^d)$ is defined as
$$V_d(\phi) := \overline{\operatorname{span}}\{\phi(\cdot - \bm{n}) : \bm{n}\in\mathbb{Z}^d\}.$$
Throughout this work, we assume that the integer translates $\{\phi(\cdot-\bm{n})\}_{\bm{n}\in\mathbb{Z}^d}$
form a Riesz basis for $V_d(\phi)$. Equivalently, there exist constants $B \ge A > 0$ such that the spectral function 
$$G_\phi(\bm{\omega}) = \sum_{\bm{n}\in\mathbb{Z}^d}|\widehat{\phi}(\bm{\omega}+\bm{n})|^2$$ satisfies
$$0 < A=\essinf_{\bm{\omega}\in [0,1]^d} G_\phi(\bm{\omega}) \leq B= \esssup_{\bm{\omega}\in [0,1]^d} G_\phi(\bm{\omega}) < \infty.$$ 
We refer to such a $\phi$ as a stable generator of $V_d(\phi)$.
Consequently, every $f\in V_d(\phi)$ admits a unique representation
\begin{align}\label{f in SIS}
    f(\mathbf{t})=\sum_{\bm{n}\in \mathbb{Z}^d} c[{\bm{n}}]\phi(\mathbf{t}-\bm{n}),\quad c\in \ell^2(\mathbb{Z}^d).
\end{align}
Derivative sampling requires additional regularity. Let $r\in\mathbb N_0$ be at least as large as every total derivative order used in the sampling model. We assume that $\phi\in C^r(\mathbb R^d)$ and
$$
\Gamma_{\boldsymbol\alpha}
:=
\sup_{\boldsymbol t\in\mathbb R^d}
\sum_{\boldsymbol n\in\mathbb Z^d}
\left|
D^{\boldsymbol\alpha}
\phi(\boldsymbol t-\boldsymbol n)
\right|^2
<\infty,
\quad
|\boldsymbol\alpha|\le r.
$$
Under these sufficient conditions, each $f\in V_d(\phi)$ has a $C^r$ representative satisfying
$$
D^{\boldsymbol\alpha}f(\boldsymbol t)
=
\sum_{\boldsymbol n\in\mathbb Z^d}
c[\boldsymbol n]\,
D^{\boldsymbol\alpha}\phi(\boldsymbol t-\boldsymbol n),
\quad
|\boldsymbol\alpha|\le r,
$$
with uniform convergence. The Cauchy–Schwarz inequality gives
$$
\left|D^{\boldsymbol\beta}f(\boldsymbol t)\right|
\le
\sqrt{\Gamma_{\boldsymbol\beta}}\,
\|c\|_{\ell^2}
\le
\sqrt{\frac{\Gamma_{\boldsymbol\beta}}{A}}\,
\|f\|.
$$
Consequently, the derivative evaluations used below are well-defined bounded linear functionals on $V_d(\phi)$.

\subsection{A multichannel sampling problem}
For signals following the SIS model, we consider a multichannel derivative-sampling framework. In particular, each channel considers a specific derivative order of the signal, followed by a set of periodic non-uniform measurements. The multichannel framework allows each channel to operate on a coarser sampling lattice, while the collection of all channels retains sufficient information for stable recovery. Here, the recovery is to determine the coefficients $\{c[\mathbf{k}]\}_{\mathbf{k}\in\mathbb Z^d}$ from the samples, and ultimately the continuous signal $f\in V_d(\phi)$. 

To elaborate on our approach, we first consider the one-dimensional (1D) signals. In Papoulis's generalized sampling framework \cite{Papouli}, multichannel sampling was considered for 1D bandlimited signals and showed that signals can be sampled at $1/M$-th of the Nyquist rate by using $M$ channels. For example, the author considered derivative sampling for $M=2$, where the signal $f(t)$ and its derivative $f^{(1)}(t)$ are sampled individually at half of the Nyquist rate and are perfectly recovered from the samples. In another example, the signal and its shifted versions are sampled at low rates, yielding a periodic sampling pattern. The key is that $M$ channels allow sampling at $1/M$-th rate per channel.

For clarity, we first describe the multichannel measurement model in the two-dimensional setting before presenting its $d$-dimensional extension. Consider a signal $f(t_1, t_2)$ belonging to the shift-invariant space (SIS) as in \eqref{f in SIS}. Let $m_{1, i}, m_{2, p} \in \mathbb{N}_0$ denote the orders of derivatives along the first and second dimensions, respectively, yielding the channel outputs
\begin{equation}
    y_{m_{1, i}, m_{2, p}}(t_1, t_2) = \frac{\partial^{m_{1,i}+m_{2,p}} f(t_1, t_2)}{\partial t_1^{m_{1, i}} \partial t_2^{m_{2,p}}}.
\end{equation}
We consider $M_1$ derivative orders along the first dimension ($0 \le i \le M_1-1$) and $M_2$ derivative orders along the second dimension ($0 \le p \le M_2-1$), amounting to a total of $M = M_1 M_2$ channels. For brevity, we map each pair $(i, p)$ to a single channel index $m = i M_2 + p \in \{0, 1, \dots, M-1\}$ using lexicographical ordering. Defining the multi-index $\bm{\alpha}_m = (m_{1, i}, m_{2, p})$, we denote the output of the $m$-th channel as
\begin{equation}
    y_m(t_1, t_2) = \mathcal{D}^{\bm{\alpha}_m} f(t_1, t_2),
\end{equation}
where $\mathcal{D}^{\bm{\alpha}_m} = \frac{\partial^{m_{1,i}+m_{2,p}} }{\partial t_1^{m_{1, i}} \partial t_2^{m_{2,p}}}$

Following the derivatives, the outputs are measured at a set of periodic, non-uniform locations in both dimensions to have
\begin{multline}\label{eq:2D_measurements}
\{y_m(\tau_{1,k_1}+n_1P_1,\tau_{2,k_2}+n_2P_2): n_i\in\mathbb Z,\\
\tau_{i,k_i}\in[0,P_i),\ k_i=1,\ldots,K_i,\ i=1,2\}.
\end{multline}
A necessary condition to recover the signal from the measurements in \eqref{eq:2D_measurements} is $KM\ge P$. Section III establishes unique and stable reconstruction under uniform lower and upper singular-value bounds on the associated polyphase matrix. The sampling pattern consists of $K = K_1 \times K_2$ uniform sampling patterns with sampling intervals $P_1$ and $P_2$ along the two dimensions. By using a shift operator $\mathcal{S}_{\boldsymbol{\tau}_k}$ the sampling is denoted as $\mathcal{S}_{-\boldsymbol{\tau}_k} \mathcal{D}^{\bm{\alpha}_m} f(n_1 P_1, n_2 P_2) $, where $\boldsymbol{\tau}_k = [\tau_{1,k_1} \,\, \tau_{2,k_2}]^{\mathrm{T}}$ with $k = (k_1-1) K_2 + k_2 \in \{1, \cdots, K\}$. Note that the delays and derivatives can be implemented as linear filters, resulting in an overall $M \times K$ channels, followed by uniform subsampling in each channel and dimension.

The aforementioned framework is extended to the $d$-dimensional signals $f(\mathbf{t}), \mathbf{t} \in \mathbb{R}^d$, where there are a total of $M_i$ derivatives and $K_i$ delays in the $i$-th dimension, and the subsampling interval is $P_i$ in that dimension. This results in an overall $M = \prod_{i=1}^dM_i$ derivative channels and $K = \prod_{i=1}^d K_i$ uniform sampling patterns. We would like to stress that, although there are $M_i$ derivatives in the $i$-th dimension, the order of the derivatives can be any $M_i$ natural numbers. As in the 2-D case, we make use of the lexicographical ordering for both derivatives and shifts, and denote the samples as 
\begin{align}\label{eq:d-samples}
y_{m,k}[\bm{n}]&=\mathcal S_{-\boldsymbol\tau_k}\mathcal D^{\boldsymbol\alpha_m}f(\mathbf P\odot\bm{n}),\\[-0.2em]
&k=1,\ldots,K,\quad m=0,\ldots,M-1.\nonumber
\end{align}
where $\bm{n} \in \mathbb{Z}^d$, $\mathbf{P} = [P_1, \dots P_d]^{\mathrm{T}}$, $\boldsymbol{\alpha}_m \in \mathbb{N}_0^d$, and $\boldsymbol{\tau}_k \in \mathbb{R}^d$. The samples in \eqref{eq:d-samples} can be considered as a result of a linear filtering followed by sampling, with the frequency response of the filter being given as
\begin{align}
    H_{m, k}(\bomega) = (2\pi \mathrm{j} \boldsymbol{\omega})^{\balpha_m}\, e^{\mathrm{2\pi j \bomega \cdot \btau_k }}. \label{eq:freq_response}
\end{align}
Using the filter's impulse response $h_{m, k}(\mathbf{t})$, the samples can be written as
\begin{align}
    y_{m, k}[\bm{n}] = (f *h_{m, k})(\mathbf{t})|_{\mathbf{t} = \mathbf{P} \odot \bm{n}}. \label{eq:filter_meas}
\end{align}

The objective is to recover $f(\mathbf{t})$ in SIS as in \eqref{f in SIS} from the measurements \eqref{eq:filter_meas}.

\section{Multidimensional Reconstruction via Polyphase Analysis}
\label{sec: 3}
The central problem is to determine whether the multichannel measurements introduced in Section \ref{sec: 2} uniquely and stably determine the underlying coefficient sequence. To address this question, we derive a finite-dimensional polyphase representation of the sampling system and characterize stable reconstruction through uniform singular-value bounds on the resulting frequency-dependent matrix.

To this end, first, consider the Fourier transform of a signal $f\in V_d(\phi)$ (cf. \eqref{f in SIS}) given as 
$$\widehat{f}(\bm{\omega}) = D(\bm{\omega})\, \widehat{\phi}(\bm{\omega}),$$
where $D(\bm{\omega}) = \sum_{\bm{n}} c[\bm{n}] \,e^{-2\pi \mathrm{j}\bm{\omega} \cdot \bm{n}}$.  
Then, by using a polyphase decomposition as in \eqref{eq:d-polyphase}, we have
\begin{align}\label{D(omega) eqn}
    D(\bm{\omega})&= \sum_{\bm{l}\in \mathcal{I}} \sum_{\bm{n}\in \mathbb{Z}^d} c[\mathbf{P\odot n}+\bm{l}]\, e^{-2\pi \mathrm{j}\bm{\omega}\cdot {(\mathbf{P\odot n}+\bm{l})}}\nonumber\\
    &=\sum_{\bm{l}\in \mathcal{I}} e^{-2\pi \mathrm{j}\bm{\omega} \cdot \bm{l}}\, C_{\bm{l}}(\bm{\omega}),
\end{align}
where 
$$C_{\bm{l}}(\bm{\omega})=\sum_{\bm{n}\in \mathbb{Z}^d} c[\mathbf{P\odot n}+\bm{l}]\, e^{-2\pi \mathrm{j}\bm{\omega}\cdot \left(\mathbf{P}\odot\bm{n}\right)},\quad\text{for}\quad \bm{l}\in \mathcal{I}.$$
Moreover, note that $C_{\bm{l}}\left(\bm{\omega}\mathbf{+\frac{n}{P}}\right)=C_{\bm{l}}(\bm{\omega})$ for $\bm{n}\in \mathbb{Z}^d$. Here, $\frac{\bm{n}}{\mathbf{P}}=\left(\frac{n_1}{P_1},\frac{n_2}{P_2},\cdots,\frac{n_d}{P_d}\right)$. 
Next, consider the measurements
\begin{align}\label{fourier of samples}
y_{m,k}[\bm{n}]=\mathcal{S}_{-\boldsymbol{\tau}_k}\mathcal{D}^{\boldsymbol{\alpha}_m}f(\mathbf{P}\odot \bm{n}),\quad\bm{n}\in\mathbb{Z}^d,
\end{align}
and its discrete-time Fourier transform is given as
$$\mathbf{Y}_{m,k}(\bm\omega)=
\sum_{\bm{n}\in\mathbb Z^d}
y_{m,k}[\bm{n}]\, e^{-2\pi \mathrm{j}(\mathbf P \odot\bm{n})\cdot\bm\omega}.$$
By using \eqref{eq:filter_meas} and applying the Poisson summation formula, we obtain
$$\mathbf{Y}_{m,k}(\boldsymbol{\omega})=\frac{1}{P}\sum_{\bm{n}\in\mathbb Z^d}
H_{m,k}\!\left(
\boldsymbol{\omega}-\frac{\bm{n}}{\mathbf P}\right)\widehat f\left(\boldsymbol{\omega}-\frac{\bm{n}}{\mathbf P}\right).$$
Using
$$ \widehat f \left( \bm\omega-\frac{\bm{n}}{\mathbf P} \right) = D \left( \bm\omega-\frac{\bm{n}}{\mathbf P} \right) \widehat\phi \left( \bm\omega-\frac{\bm{n}}{\mathbf P} \right) $$
together with \eqref{D(omega) eqn} and the periodicity of the functions $C_{\bm l}$, we obtain
\begin{align}\label{E:DTFT of y_m}
\mathbf{Y}_{m,k}(\boldsymbol{\omega})&=\sum_{\bm l\in\mathcal I}
C_{\bm l}(\boldsymbol{\omega})
e^{-2\pi \mathrm{j}\boldsymbol{\omega}\cdot\bm l}\,\Phi^{\bm l}_{m,k}
(\boldsymbol{\omega}),
\end{align}
where
\begin{align}\label{derivative}
\Phi^{\bm l}_{m,k}(\boldsymbol{\omega})=\frac1{P}\sum_{\bm{n}\in\mathbb Z^d}H_{m,k}\!\left(\boldsymbol{\omega}-\frac{\bm{n}}{\mathbf P}\right)\,e^{2\pi \mathrm{j}\frac{\bm{n}}{\mathbf P}\cdot\bm l}\,\widehat\phi\!\left(\boldsymbol{\omega}-\frac{\bm{n}}{\mathbf P}\right)
\end{align}
for $m\in \{0,1,\dots, M-1\}$ and $ k\in \{1,2,\dots, K\}$.
We collect the $P$ polyphase components and the $K\!M$ measurement sequences into the column vectors
 $$\widetilde{\mathbf{C}}(\bm{\omega})=\big[C_{\bm{l}}(\bm \omega)\, e^{-2\pi\mathrm{j}\bm\omega\cdot\bm{l}}\big]^{\mathrm T}_{\bm{l}\in \mathcal{I}}\quad \text{and}\quad\mathbf{Y}(\bm{\omega})=\big[\mathbf{Y}_{m,k}(\bm\omega)\big]^{\mathrm T},$$
 respectively. Then, \eqref{E:DTFT of y_m} can be written as the finite-dimensional system
 $$\mathbf{Y}(\bm{\omega}) = \mathbf{\Phi}(\bm{\omega})\, \widetilde{\mathbf{C}}(\bm{\omega}),$$
 where $\mathbf\Phi(\boldsymbol{\omega})$ is a $K\!M\times P$ polyphase matrix with entries $
\Phi^{\bm l}_{m,k}(\boldsymbol{\omega})$ which is defined in \eqref{derivative}.

Suppose that the polyphase matrix satisfies the uniform bounds
$$\essinf_{\bm\omega}
\sigma_{\min}(\Phi(\bm\omega))>0
\text{ and }\esssup_{\bm\omega}
\sigma_{\max}(\Phi(\bm\omega))<\infty,$$
which implies $\rank \mathbf\Phi(\boldsymbol{\omega})=P$ for almost every $\bm\omega$ with a necessary condition $K\!M\ge P$. Then $\mathbf\Phi(\boldsymbol{\omega})$ has full column rank, and its Moore-Penrose left inverse is
$$\mathbf\Phi^\dagger(\boldsymbol{\omega})=\left(\mathbf\Phi^*(\boldsymbol{\omega})\mathbf\Phi(\boldsymbol{\omega})\right)^{-1}\mathbf\Phi^*(\boldsymbol{\omega}).$$
Hence, $\widetilde{\mathbf{C}}(\bm{\omega})=\mathbf\Phi^\dagger(\boldsymbol{\omega})\,\mathbf{Y}(\bm{\omega})$ and combining it with \eqref{fourier of samples}, results in
\begin{align*}
C_{\bm l}(\bm\omega)e^{-2\pi\mathrm j\bm\omega\cdot\bm l}
&=\sum_{m=0}^{M-1}\sum_{k=1}^{K}[\Phi^\dagger(\bm\omega)]^{\bm l}_{m,k}Y_{m,k}(\bm\omega)\\
&=\sum_{m=0}^{M-1}\sum_{k=1}^{K}\sum_{\bm{n}\in\mathbb Z^d}
\mathcal D^{\balpha_m}f(\btau_k+\mathbf P\odot\bm{n})\\[-0.2em]
&\quad\times[\Phi^\dagger(\bm\omega)]^{\bm l}_{m,k}
 e^{-2\pi\mathrm j(\mathbf P\odot\bm{n})\cdot\bm\omega}.
\end{align*}
By using \eqref{D(omega) eqn}, we obtain the DTFT of the coefficients from the samples as
\begin{align}
D(\bm\omega)=&\sum_{m=0}^{M-1}\sum_{k=1}^{K}\sum_{\bm{n}\in\mathbb Z^d}
\mathcal D^{\boldsymbol\alpha_m}f(\btau_k+\mathbf P\odot\bm{n})\nonumber\\[-0.2em]
&\quad\times e^{-2\pi\mathrm j(\mathbf P\odot\bm{n})\cdot\bm\omega}
\sum_{\bm l\in\mathcal I}[\Phi^\dagger(\bm\omega)]^{\bm l}_{m,k}.
\end{align}
Now, multiplying $D(\omega)$ by $\widehat{\phi}(\omega)$ and then taking the inverse Fourier transform, we obtain the reconstruction formula
$$f_{\text{rec}}(\mathbf{t}) =\sum_{m=0}^{M-1}\sum_{k=1}^K \, \sum_{\bm{n}\in\mathbb{Z}^d} D^{\boldsymbol{\alpha}_m}f(\btau_k+\mathbf P \odot \bm{n})\,\psi_{m,k}(\mathbf{t} -\mathbf P\odot\bm{n})$$
where the \textbf{synthesis kernels}, $\psi_{m,k}(\mathbf{t})$ are defined by
$$\widehat{\psi}_{m,k}(\bm{\omega}) =\widehat{\phi}(\bm{\omega}) \sum_{\bm{l}\in \mathcal{I}} [\mathbf{\Phi}^\dagger(\bm{\omega})]^{\bm{l}}_{m,k}.$$

\begin{figure*}[!t]
\centering
\includegraphics[width=7in]
{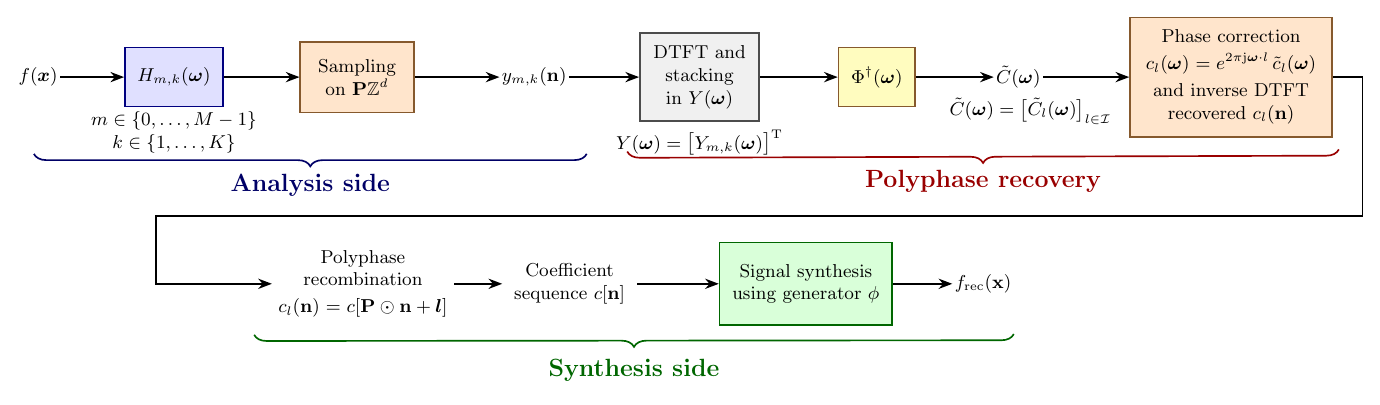}
\caption{Multichannel sampling and reconstruction system with polyphase
matrix. Each analysis channel corresponds to a prescribed derivative
order and periodic sampling shift, and its output is sampled on the coarse lattice $\mathbf P\mathbb Z^d$.}
\label{fig:Figure1}
\end{figure*}
We now state the main reconstruction theorem.  Uniform lower and upper singular-value bounds on the polyphase matrix $\Phi(\bm\omega)$ guarantee unique and stable reconstruction of every function in $V_d(\phi)$. In this case, the synthesis kernels are obtained from the Moore-Penrose pseudoinverse of $\mathbf{\Phi}(\bm{\omega})$. The complete analysis and reconstruction pipeline, including multichannel sampling, polyphase recovery, coefficient recombination, and signal synthesis, is summarized in Figure \ref{fig:Figure1}.
\begin{thm}\label{main_thm}
Let $\phi\in L^2(\mathbb R^d)$ be a stable generator of $V_d(\phi)$, and suppose that $K\! M\ge P$. Let $\Omega_{\mathbf P}:=\prod_{i=1}^{d}[0,1/P_i)$. If the associated polyphase matrix $\Phi(\bm\omega)\in\mathbb C^{KM\times P}$ satisfies
\begin{align*}
\essinf_{\bm\omega\in\Omega_{\mathbf P}}\sigma_{\min}(\Phi(\bm\omega))>0,~\text{ and }~\esssup_{\bm\omega\in\Omega_{\mathbf P}}\sigma_{\max}(\Phi(\bm\omega))<\infty,
\end{align*}
then every $f\in V_d(\phi)$ is uniquely and stably determined by
\begin{multline*}
\{\mathcal D^{\boldsymbol\alpha_m}f(\btau_k+\mathbf P\odot\bm{n}):\ m=0,\ldots,M-1,\\
k=1,\ldots,K,\ \bm{n}\in\mathbb Z^d\}.
\end{multline*}
Moreover,
\begin{align}\label{sampling_formula}
f_{\rm rec}(\mathbf t)=\sum_{m=0}^{M-1}\sum_{k=1}^{K}\sum_{\bm{n}\in\mathbb Z^d}
&\mathcal D^{\boldsymbol\alpha_m}f(\btau_k+\mathbf P\odot\bm{n})\nonumber\\[-0.2em]
&\times\psi_{m,k}(\mathbf t-\mathbf P\odot\bm{n}).
\end{align}
where the synthesis kernels $\psi_{m,k}$ are determined by the Moore--Penrose pseudoinverse $\mathbf\Phi^\dagger(\boldsymbol{\omega})$.
\end{thm}

\begin{rem}
If $K\!M=P$, then $\Phi(\boldsymbol\omega)$ is square. If it is uniformly invertible on the fundamental frequency domain, then the Moore–Penrose left inverse reduces to the ordinary inverse $\Phi^{-1}(\boldsymbol\omega)$. In this case, the synthesis kernels, $\psi_{m,k}(\mathbf{t})$ are defined by
$$\widehat{\psi}_{m,k}(\bm{\omega}) =\widehat{\phi}(\bm{\omega}) \sum_{\bm{l}\in \mathcal{I}} [\mathbf{\Phi}^{-1}(\bm{\omega})]^{\bm{l}}_{m,k},$$
for $m\in\{0,1,\dots, M-1\}$ and $k\in\{1,\dots,K\}$.
\end{rem}

\begin{rem}
Let $a,b>0$ be uniform lower and upper singular-value bounds for $\Phi$, and let $A, B$ be the Riesz bounds of $\varphi$. Then
\[
\frac{a^2}{B}\|f\|^2
\le
\sum_{m=0}^{M-1}\sum_{k=1}^{K}
\|y_{m,k}\|_{\ell^2}^2
\le
\frac{b^2}{A}\|f\|^2.
\]
These inequalities follow from Parseval’s identity and the Riesz bounds.
\end{rem}

The following examples illustrate how the general formulation recovers classical one-dimensional sampling formulas and extends to non-separable multidimensional generators.

\subsection{Illustrative Examples}
In each numerical example, the exactness of the explicit reconstruction formula is validated by computing the relative recovery error using $\frac{\|f-f_{\text{rec}}\|}{\|f\|}$.
\begin{example} (\textbf{1D Derivative Sampling})
Let $\phi(t)=\sinc t$, $P=2$, and $M=2$. From \eqref{derivative},
\begin{align*}
\Phi(\omega)&=\begin{bmatrix}1&0\\ \frac{\mathrm j\pi}{2}(4\omega-1)&\frac{\mathrm j\pi}{2}\end{bmatrix},\\[-0.2em]
\Phi(\omega)^{-1}&=\begin{bmatrix}1&0\\ 1-4\omega&-\frac{2\mathrm j}{\pi}\end{bmatrix}.
\end{align*}
Hence $\psi_0(t)=\sinc^2(t/2)$ and $\psi_1(t)=t\sinc^2(t/2)$, giving
\begin{align*}
f_{\rm rec}(t)=\sum_{n\in\mathbb Z}&\Big[f(2n)\sinc^2\!\left(\frac{t-2n}{2}\right)\\
&+f'(2n)(t-2n)\sinc^2\!\left(\frac{t-2n}{2}\right)\Big].
\end{align*}
This is the classical derivative-sampling formula in \cite{Papouli} for $\sigma=\pi$.
\end{example}

\begin{example}
Let $\phi(t)=\sinc t$ and $\Lambda=(\tau_1+2\mathbb Z)\cup(\tau_2+2\mathbb Z)$, $\tau_1\neq\tau_2$. With $H_{0,k}(\omega)=e^{2\pi\mathrm j\tau_k\omega}$, the two $2\times2$ polyphase matrices on $[0,1/2]$ and $[-1/2,0)$ have determinants
$$
\tfrac12 e^{2\pi\mathrm j(\tau_1+\tau_2)\omega}(e^{\mp\pi\mathrm j\tau_1}-e^{\mp\pi\mathrm j\tau_2})\neq0,
$$
respectively. Inverting them and taking the inverse Fourier transform yields
\begin{align*}
\psi_0(t)&=\frac{\sin\!\left(\frac\pi2(t-\tau_2)\right)}{\sin\!\left(\frac\pi2(\tau_1-\tau_2)\right)}\sinc\!\left(\frac{t-\tau_1}{2}\right),\\
\psi_1(t)&=\frac{\sin\!\left(\frac\pi2(t-\tau_1)\right)}{\sin\!\left(\frac\pi2(\tau_2-\tau_1)\right)}\sinc\!\left(\frac{t-\tau_2}{2}\right).
\end{align*}
The resulting reconstruction is the periodic nonuniform formula reported in \cite{Papouli} for $-\tau_1=\tau_2=\tau$.
\end{example}

\subsection{Multidimensional Separable and Non-Separable Generators}
\begin{example}\label{Example 3.4}
    Let $\phi(t_1,t_2)=\sinc{(t_1)}\sinc{(t_2)}$, $M_1=1$, $M_2=2$ and $\mathbf P=(1,2)$. We take
    $$H_{0,1}(\omega_1,\omega_2)=1\text{ and }H_{1,1}(\omega_1,\omega_2)=2\pi \mathrm{j}\omega_2.$$
    We obtain from \eqref{derivative} that
\begin{align*}
    \mathbf{\Phi}(\omega_1,\omega_2)=\frac{1}{2}\begin{bmatrix}
    2&0\\[0.6em]
    \mathrm{j}\pi(4\omega_2-1)& \mathrm{j}\pi
    \end{bmatrix}.
\end{align*}
Hence, we obtain the following sampling formula
\begin{align*}
f_{\rm rec}(t_1,t_2)=4\sin^2\!\frac{\pi t_2}{2}\sum_{m,n\in\mathbb Z}
&\left[\frac{f(m,2n)}{\pi^2(t_2-2n)^2}
+\frac{\partial_{t_2}f(m,2n)}{\pi^2(t_2-2n)}\right]\\[-0.2em]
&\times\frac{\sin\pi(t_1-m)}{\pi(t_1-m)}.
\end{align*}
\end{example}

\begin{figure}[!t]
\centering
\includegraphics[width=\columnwidth]{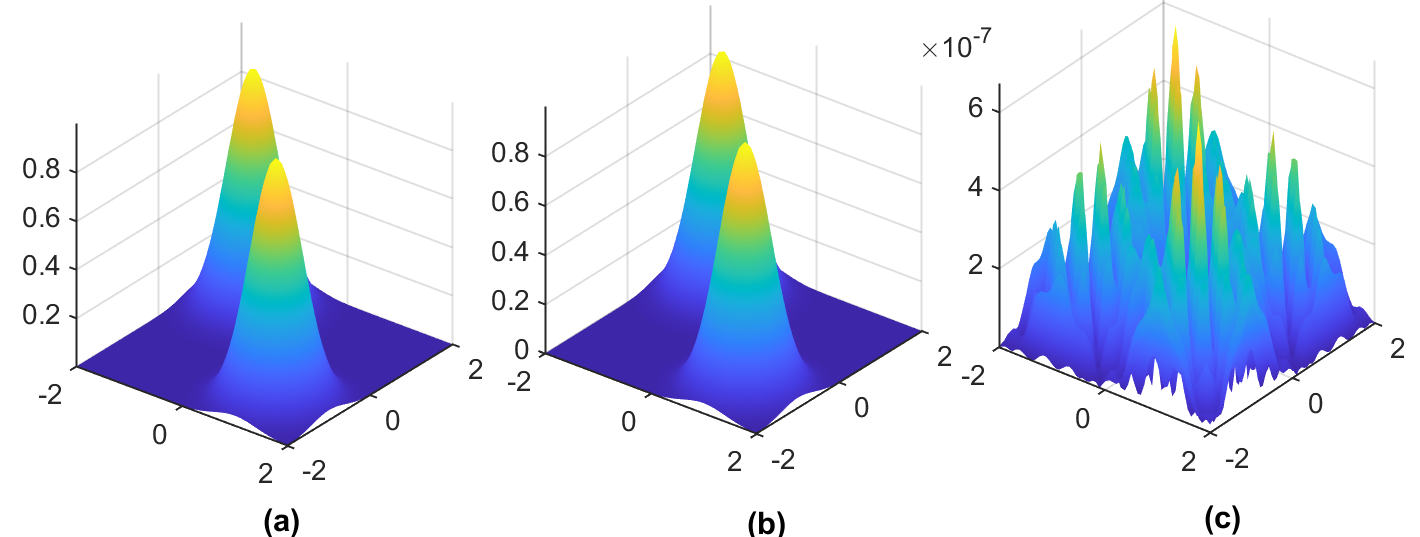}
\caption{(a) The original function $f(t_1,t_2)=\phi(t_1-1,t_2+1)+\phi(t_1+1,t_2-1)$, (b) the reconstructed function, and $(c)$ the absolute error. The relative error is $7.202266\times 10^{-7}$.}
\label{figure 3}
\end{figure}

Sampling in Gaussian-generated shift-invariant spaces, in the setting of function-value samples rather than derivative measurements, has been studied in \cite{romero2024sampling, pfander2013remarks}. In this paper, we consider derivative sampling.
\begin{example}\label{Example 3.5}
    Let $\phi(t_1,t_2)=e^{-\pi(t_1^2+t_2^2)}$ and  $M_1=2$, $M_2=1$, $\mathbf P=(2,1)$ with the sampling sets $\Lambda_1=\frac{1}{4}+2\mathbb{Z}$ along $x$-axis and $\Lambda_2=\left(\frac{1}{3}+\mathbb{Z}\right)\cup\left(\frac{2}{3}+\mathbb{Z}\right)$ along $y$-axis.
The corresponding analysis filters are
$$ H_{0,k}(\omega_1,\omega_2) = e^{2\pi \mathrm{j}\left(\frac{1}{4}\omega_1+\tau_{2,k_2}\omega_2\right)}$$
and
$$ H_{1,k}(\omega_1,\omega_2) = 2\pi \mathrm{j}\omega_1 e^{2\pi \mathrm{j}\left(\frac{1}{4}\omega_1+\tau_{2,k_2}\omega_2\right)}$$
where $ \tau_{2,1}=\frac13$ and $\tau_{2,2}=\frac23$, and $k=k_2$ with $k_2=1,2$.
Since $\mathcal I=\{(0,0),(1,0)\}, $
the associated polyphase matrix $\mathbf{\Phi}(\omega_1,\omega_2)$ is a $4\times2$ matrix on the fundamental domain $(\omega_1,\omega_2)\in [0,\tfrac12)\times[0,1).$ Its entries are
\begin{align*}
\Phi_{m,k}^{\bm l}(\omega_1,\omega_2)=\frac12\sum_{(n_1,n_2)\in\mathbb Z^2}
&\left(2\pi\mathrm j\left(\omega_1-\frac{n_1}{2}\right)\right)^m e^{\pi\mathrm j n_1l_1}\\[-0.2em]
&\times e^{2\pi\mathrm j[\frac14(\omega_1-n_1/2)+\tau_{2,k_2}(\omega_2-n_2)]}\\[-0.2em]
&\times e^{-\pi[(\omega_1-n_1/2)^2+(\omega_2-n_2)^2]}.
\end{align*}
for $\bm l=(l_1,l_2)\in \mathcal{I}$ and $m=0,1$. Moreover, we can rewrite the above entries as 
$$\Phi_{m,k}^{(l_1,0)}(\omega_1,\omega_2)=\frac{1}{2} Y_k(\omega_2) X_m^{l_1}(\omega_1),$$
where
$$Y_k(\omega_2) = \sum_{n\in\mathbb{Z}} e^{2\pi \mathrm{j}\, \tau_{2,k_2}(\omega_2-n)} e^{-\pi(\omega_2-n)^2},\quad k_2=1,2$$
and $X_m^l(\omega_1)$
\begin{align*}
=\sum_{n\in\mathbb Z}\left(2\pi\mathrm j\left(\omega_1-\frac n2\right)\right)^m e^{\pi\mathrm jnl} e^{\frac{\pi\mathrm j}{2}(\omega_1-n/2)}e^{-\pi(\omega_1-n/2)^2}.
\end{align*}
for $m=0,1$ and $l=0,1$
It is therefore enough to consider the $2\times2$ minor obtained from the two rows corresponding to $\tau_{2,1}=1/3$, namely
\begin{align*}
\Phi_1(\omega_1,\omega_2)& = \frac{Y_1(\omega_2)}{2} 
    \begin{pmatrix} 
    X_0^0(\omega_1) & X_0^1(\omega_1)\\[0.2em] 
    X_1^0(\omega_1) & X_1^1(\omega_1) 
\end{pmatrix}\\
&:=\frac{Y_1(\omega_2)}{2} A(\omega_1).
\end{align*}
By Poisson summation,
$$ Y_1(\omega_2) = \sum_{n\in\mathbb Z} e^{-\pi(n-\frac13)^2}e^{2\pi \mathrm{j} n\omega_2}. $$
Hence
$$ |Y_1(\omega_2)| \ge e^{-\pi/9} - \sum_{n\neq0}e^{-\pi(n-\frac13)^2} >0.45, $$
uniformly in $\omega_2$. Similarly, the determinant of $A(\omega_1)$ can be written as
\begin{align*}
\det A(\omega_1)=8\pi e^{2\pi\mathrm j\omega_1}\sum_{i,n\in\mathbb Z}&(2(i-n)+1)\\[-0.2em]
&\times e^{-\pi[(2n-\frac14)^2+(2i+\frac34)^2]}e^{4\pi\mathrm j(i+n)\omega_1}.
\end{align*}
The term $i=n=0$ has modulus $e^{-5\pi/8}$, while the sum of the absolute values of all remaining terms is less than $0.019$. Therefore,
$$ |\det A(\omega_1)| > 8\pi\left(e^{-5\pi/8}-0.019\right)>0 $$
for every $\omega_1\in[0,1/2]$. Consequently,
$\det\Phi_1(\omega_1,\omega_2)\neq0 $
throughout the fundamental domain. It suffices to consider the two rows corresponding to $\tau_{2,1}=1/3$. Indeed, since $\Phi_1$ denotes this $2\times2$ submatrix, then
$$ \Phi(\omega)^*\Phi(\omega)\succeq \Phi_1(\omega)^*\Phi_1(\omega), $$
and therefore
$$ \rank\bm \Phi(\omega_1,\omega_2)=2. $$
Moreover,
$$\sigma_{\min}(\mathbf \Phi(\bm\omega))\ge \sigma_{\min}(\Phi_1(\bm\omega)).$$
Since singular values of $\bm\Phi$ are continuous and periodic, they define continuous functions on the compact frequency torus, which may equivalently be represented by $[0,\frac12]\times[0,1]$. Hence
$$ \min_{\omega}\sigma_{\min}(\mathbf \Phi(\bm\omega))>0, \quad \max_{\omega}\sigma_{\max}(\mathbf \Phi(\bm\omega))<\infty. $$
Thus the uniform stability condition of Theorem \ref{main_thm} is satisfied. The corresponding reconstruction obtained from the Moore--Penrose pseudoinverse is illustrated in Figure \ref{figure 3}.
\end{example}

Tensor-product sampling constructions exploit separability of the generator to organize reconstruction along individual coordinates. Our polyphase formulation represents all measurement channels jointly through a multidimensional matrix system. For separable generators and product sampling patterns, it recovers the corresponding tensor-product structure. The same formulation also applies to non-separable generators without requiring such a factorization.

\begin{figure}[!t]
\centering
\includegraphics[width=\columnwidth]{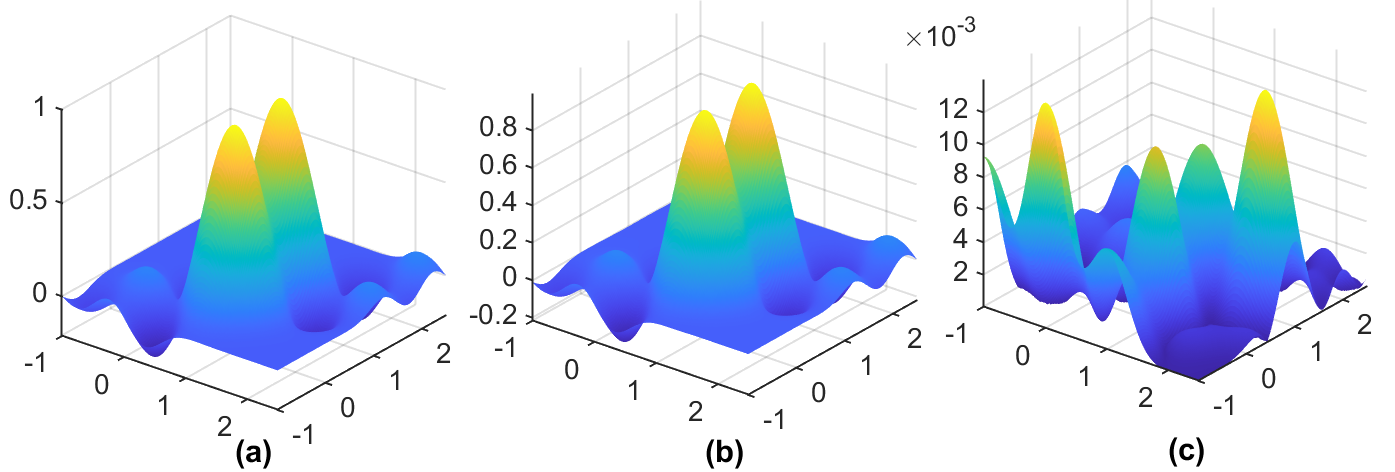}
\caption{(a) The original function $f(t_1,t_2)=\phi(t_1-1,t_2-1)+\phi(t_1-1,t_2)$, $(b)$ the reconstructed function $f_{\text{rec}}$ (middle) and (c) the absolute error $|f-f_{rec}|$.}
\label{fig:non-separable1}
\end{figure}

\begin{example}\label{Example 3.6}
Consider the non-separable, square-integrable function
$$\phi(t_1,t_2) = \text{sinc}(t_1+t_2)\,e^{-\pi(t_1-t_2)^2}.$$
We take $M_1=M_2=2$, and the sampling sets $\Lambda_1=2\mathbb{Z}$, and
 $\Lambda_2=\left(\frac{1}{5}+2\mathbb{Z}\right)\cup\left(\frac{1}{4}+2\mathbb{Z}\right)$. We define eight analysis filters
\begin{align*}
H_{m,k}(\omega_1,\omega_2)&=(2\pi\mathrm j\omega_1)^{m_1}(2\pi\mathrm j\omega_2)^{m_2}e^{2\pi\mathrm j\tau_{2,k_2}\omega_2},\\[-0.2em]
m&=2m_1+m_2,\quad k=k_2.
\end{align*}
where $0\le m_1,m_2\le 1$, $k_2=1,2$ and $\tau_{2,1}=1/5$, $\tau_{2,2}=1/4$. The Fourier transform of the generator
$$\widehat{\phi}(\omega_1, \omega_2) = \frac{1}{2}\,\chi_{[-1, 1]}\left(\omega_1+\omega_2\right)\, e^{-\frac{\pi}{4} \left({\omega_1-\omega_2}\right)^2}.$$
Moreover, 
$$G_\phi(\boldsymbol\omega)
\ge |\widehat{\phi}(\boldsymbol\omega)|^2\ge \frac14e^{-\pi/2}>0$$
almost everywhere on $[-\tfrac12,\tfrac12)^2$. The strip support restricts the admissible sums of the alias indices, while Gaussian decay in their differences gives a uniform upper bound. Hence, $\phi$ is a stable generator of $V_2(\phi)$.

For any $(\omega_1, \omega_2) \in [-1/2, 1/2)^2$, the sampling matrix $\mathbf{\Phi}(\omega_1, \omega_2) \in \mathbb{C}^{8 \times 4}$. Its rows are indexed by $(m_1,m_2,k_2)\in\{0,1\}^2\times\{1,2\}$, and its columns by $(l_1,l_2)\in\{0,1\}^2$. The entries are
\begin{align}\label{entries of Phi_3.6}
\Phi_{m,k}^{\bm l}(\omega_1,\omega_2)=&\frac14\sum_{(n_1,n_2)\in R}
H_{m,k}\!\left(\omega_1-\frac{n_1}{2},\omega_2-\frac{n_2}{2}\right)\nonumber\\[-0.2em]
&\times e^{\pi\mathrm j(n_1l_1+n_2l_2)}
\widehat\phi\!\left(\omega_1-\frac{n_1}{2},\omega_2-\frac{n_2}{2}\right).
\end{align}
where 
\begin{align*}
    R=\big\{(n_1,n_2)\in \mathbb{Z}^2:~-4\le n_1+n_2\le 4\big\}.
\end{align*}
For numerical evaluation, the infinite aliasing sum was truncated to
$ |n_1|, |n_2|\le 10$. Then the polyphase matrix over the fundamental domain gives
\begin{align*}
\min_{(\omega_1,\omega_2)\in[-1/2,1/2)^2}
\lambda_{\min}
\bigl(\Phi(\omega_1,\omega_2)^*
\Phi(\omega_1,\omega_2)\bigr)\approx 0.00915.
\end{align*}
Consequently, the polyphase matrix is numerically bounded away from singularity, in agreement with the stability condition of Theorem \ref{main_thm}. The corresponding numerical reconstruction in Figure \ref{fig:non-separable1} has relative error $0.0157$.
\end{example}

\begin{example}\label{Example 3.7}
We define the generator as a hypergeometric function
$$\phi(\mathbf{t}) = {}_2F_1(a, b; c; -\|\mathbf{t}\|^2), \quad \mathbf{t} = (t_1, t_2).$$
Since $\|\mathbf{t}\|^2 = t_1^2 + t_2^2$ cannot be linearly factored out of the hypergeometric series, $\phi(\mathbf{x})$ is strictly non-separable. For large $\|\mathbf{t}\|$, the asymptotic decay of this function is dominated by $\mathcal{O}(\|\mathbf{t}\|^{-2a}) + \mathcal{O}(\|\mathbf{t}\|^{-2b})$. Consequently,
$$|\phi(\mathbf{t})|
\le
C(1+\|\mathbf{t}\|)^{-2\min\{a,b\}}.$$
If $a,b>1$, then $\phi\in L^1(\mathbb{R}^2)\cap L^2(\mathbb{R}^2)$. Let $a=2, b=2$, and $c=1$. For more details on hypergeometric functions, see \cite{AndrewsAskeyRoy, Rainville}. 

We consider the non-separable generator 
$$\phi(t_1, t_2) = {}_2F_1(2, 2; 1; -(t_1^2 + t_2^2))=\frac{1-t_1^2-t_2^2}{(1+t_1^2+t_2^2)^3}.$$
Therefore,
$$\widehat{\phi}(\bm{\omega}) = 2\pi^3\Vert{}\bm{\omega}\Vert{}^2 B_0(2\pi\Vert{}\bm{\omega}\Vert{}),$$
where $B_0$ denotes the modified Bessel function of the second kind of order zero, and the value at $\bm{\omega}=0$ is understood by continuity as $\widehat{\phi}(0)=0$. Since $B_0(t)>0$ for $t>0$, we have $\widehat{\phi}(\bm{\omega})>0$ for $\bm{\omega} \neq 0$.
Furthermore, 
$$G_\phi(\bm{\omega}) = \sum_{\mathbf{k}\in\mathbb{Z}^2} \left\vert{} \widehat{\phi}(\bm{\omega}+\mathbf{k}) \right\vert{}^2$$
is continuous and strictly positive on $[0,1]^2$. The exponential decay of $B_0$ also implies that $G_\phi$ is bounded above. Consequently,
$$0 < \inf_{\bm{\omega}\in[0,1]^2}G_\phi(\bm{\omega}) \le \sup_{\bm{\omega}\in[0,1]^2}G_\phi(\bm{\omega}) < \infty,$$
and therefore $\phi$ is a stable generator of $V_2(\phi)$.

Now take $\mathbf{P}=(2,2)$ and consider the sampling sets on
$\Lambda_1 = \left(\frac{1}{5}+2\mathbb{Z}\right) \cup \left(\frac{2}{5}+2\mathbb{Z}\right), \text{ and } \Lambda_2 = \frac{1}{4}+2\mathbb{Z}.$
For $k_1=1,2$, $k_2=1$ and $0\le m_1,m_2\le 1$, define the analysis filters by 
$$H_{m,k}\left({\omega }_{1},\; {\omega }_{2}\right)={\left(2\pi \mathrm{j}{\omega }_{1}\right)}^{m_{1}}{\left(2\pi \mathrm{j}{\omega }_{2}\right)}^{m_{2}}e^{2\pi \mathrm{j}\left(\tau_{1,k_1}\,{\omega }_{1}+\frac{\omega_{2}}{4}\right)},$$
where $m=2m_1+m_2$ and $k=k_1$.
Further, we have 
\begin{align*}
\partial_{t_i}{}_2F_1(a,b;c;-\|\mathbf t\|^2)
={}&-2t_i\frac{ab}{c}\\[-0.2em]
&\times{}_2F_1(a+1,b+1;c+1;-\|\mathbf t\|^2).
\end{align*}
Thus, the measurements consist of the function values and the derivatives $f, f_{t_1}, f_{t_2}, f_{t_1t_2}$ at the two shifted sampling lattices
$\left(\tau_{1,k_1}+2n_1,\frac{1}{4}+2n_2\right), \quad k_1=1,2,\quad n_1,n_2\in\mathbb{Z}.$
The entries of the corresponding polyphase matrix $\Phi(\omega_1,\omega_2) \in \mathbb{C}^{8\times4}$ are
\begin{align*}
\Phi_{m,k}^{\bm l}(\omega_1,\omega_2)&=\frac14\sum_{(n_1,n_2)\in\mathbb Z^2}
H_{m,k}\!\left(\omega_1-\frac{n_1}{2},\omega_2-\frac{n_2}{2}\right)\\[-0.2em]
&\times e^{\pi\mathrm j(n_1l_1+n_2l_2)}
\widehat\phi\!\left(\omega_1-\frac{n_1}{2},\omega_2-\frac{n_2}{2}\right).
\end{align*}
for $\bm l=(l_1,l_2)$. Because $\widehat{\phi}$ decays exponentially, the series converges rapidly, and the entries of $\Phi$ are continuous on the fundamental frequency domain. A numerical evaluation over $[-1/2,1/2)^2$ gives
$$\min_{\bm{\omega}} \lambda_{\min} \left( \Phi(\bm{\omega})^* \Phi(\bm{\omega}) \right) \approx 0.027.$$
\begin{figure}[!t]
\centering
\includegraphics[width=\columnwidth]{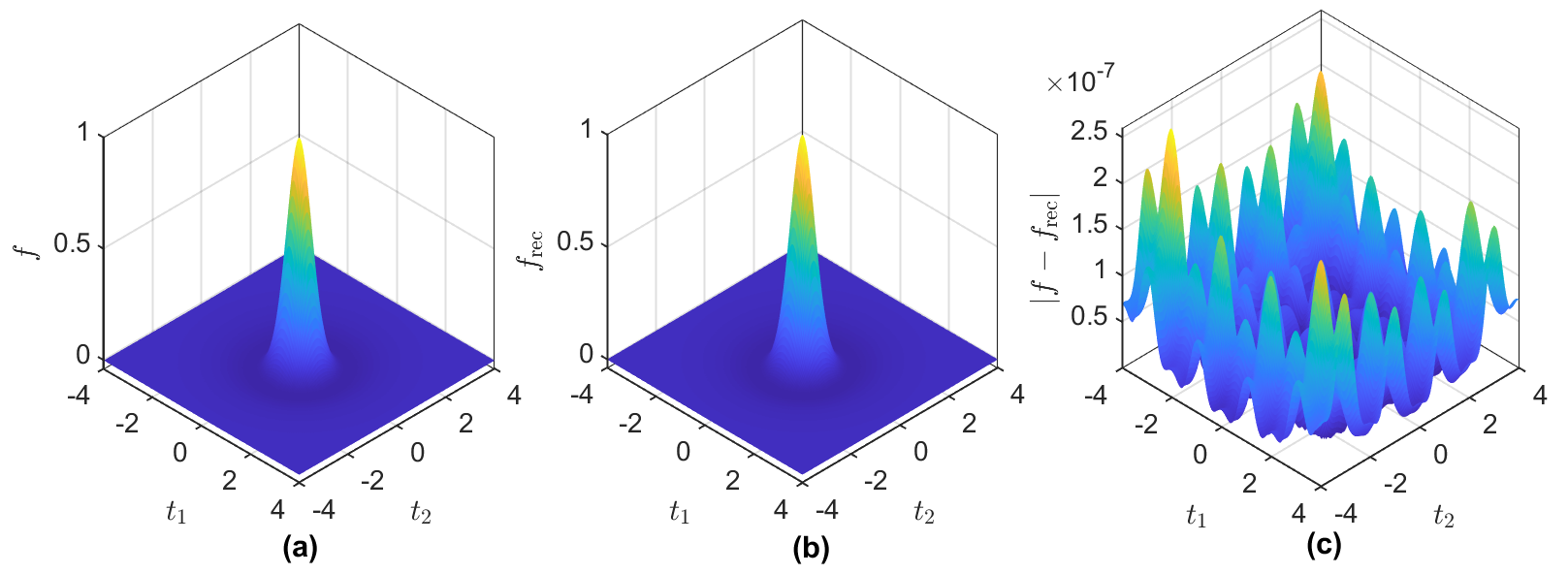}
\caption{(a) The original function $f(\mathbf{t})=\phi(\mathbf{t})$ (left), (b) the reconstructed function $f_{\text{rec}}$ (middle), and (c) the absolute error $|f-f_{rec}|$.}
\label{fig:non_separable2}
\end{figure}
Equivalently, $\min_{\bm{\omega}} \sigma_{\min}(\Phi(\bm{\omega})) \approx 0.164$.
Thus, the polyphase matrix is numerically well-separated from singularity, satisfying the stability condition. Every $f \in V_2(\phi)$ can be reconstructed using the synthesis kernels determined by the Moore-Penrose pseudoinverse $\Phi^\dagger(\bm{\omega})$. For this example, the numerical reconstruction in Figure \ref{fig:non_separable2} has relative error $8.268\times10^{-7}$.
\end{example}

\begin{rem}
The nonzero reconstruction errors in Examples \ref{Example 3.4}--\ref{Example 3.7} arise from finite truncation, discretization, numerical quadrature/FFT approximation, and floating-point arithmetic; the underlying reconstruction identities are exact whenever the hypotheses of Theorem \ref{main_thm} hold.
\end{rem}

\section{Iterative Reconstruction from Irregular Derivative Samples}
The reconstruction formulas of Section III apply to structured periodic sampling. For irregular locations, we instead construct a local bivariate Hermite interpolant from the available derivative measurements and project it onto $V_2(\phi)$. This complements earlier irregular multivariate sampling methods \cite{AldroubiFeichtinger1998,Sun2005}, multivariate Hermite interpolation \cite{Chawla}, and one-dimensional derivative reconstruction in SISs \cite{Priyanka_Selvan_JCAM}. We restrict the analysis to $\mathbf t=(x,y)$.

Let $X=\{\xi_i:i\in\mathbb Z\}$ and $Y=\{\eta_j:j\in\mathbb Z\}$ be strictly increasing separated sequences and, for $m,n\ge1$, set
$$
\delta_X^m=\sup_i(\xi_{i+m}-\xi_i),\quad \delta_Y^n=\sup_j(\eta_{j+n}-\eta_j).
$$
We reconstruct $f\in V_2(\phi)$ from
$$
\Lambda=\{\mathcal D^{(k,l)}f(\xi_i,\eta_j):0\le k\le r,\ 0\le l\le s\}.
$$
The Hermite interpolation/projection composition defines an approximation operator; under the density condition derived below, its error operator is contractive and yields geometric convergence.

\subsection{Local Hermite Interpolation} 
Let $C^{r+s}([a,b]\times[c,d])$ denote the class of functions whose mixed partial derivatives of total order up to $r+s$ are continuous on $[a,b]\times[c,d]$. Let us define
$$ R_x=m(r+1)+r,\text{ and } R_y=n(s+1)+s.$$
The quantities $R_x$ and $R_y$ arise from the number of interpolation conditions. In the $x$-direction, $m+1$ nodes are used and derivatives of orders $0,\ldots,r$ are prescribed at every node, giving $(m+1)(r+1)=R_x+1$ conditions. The same reasoning gives $R_y+1$ conditions in the $y$-direction.

\begin{lem}[\bf Hermite interpolation]\cite{Chawla}\label{Hermite_chawla}
Let $f\in C^{r+s}([a,b]\times[c,d])$, let $\xi_i\in[a,b]$, $0\le i\le m$, and $\eta_j\in[c,d]$, $0\le j\le n$, be distinct nodes.
Then there exists a unique polynomial $H_{R_x,R_y}(x,y)$, of degree at most $R_x$ in $x$ and at most $R_y$ in $y$, satisfying
$$ \mathcal D^{(k,l)} H_{R_x,R_y}(\xi_i,\eta_j) = \mathcal D^{(k,l)}f(\xi_i,\eta_j), $$
for $0\le i\le m$, $0\le j\le n$, $0\le k\le r$, and $0\le l\le s$.
\end{lem}

For each $\nu\in\mathbb Z$, let $\mathcal I_{X,\nu}$ denote the local Hermite interpolation operator in the $x$-variable constructed on
$J_\nu^X=[\xi_{m\nu},\xi_{m\nu+m}]$
using the $m+1$ consecutive nodes
$\xi_{m\nu},\xi_{m\nu+1},\ldots,\xi_{m\nu+m}$
and the derivative data
$$ \partial_x^k f(\xi_{m\nu+a},y), \text{ for } 0\le a\le m,\quad 0\le k\le r. $$
Define $\mathcal I_{Y,\mu}$ analogously on $J_\mu^Y=[\eta_{n\mu},\eta_{n\mu+n}]$
using $n+1$ consecutive $y$-nodes and derivatives up to order $s$.

The corresponding global piecewise interpolation operators $\mathcal I_X$ and $\mathcal I_Y$ are defined by
$$(\mathcal I_X f)(x,y)=(\mathcal I_{X,\nu}f)(x,y),
\quad x\in J_\nu^X,$$
and
$$(\mathcal I_Y f)(x,y)=(\mathcal I_{Y,\mu}f)(x,y),
\quad y\in J_\mu^Y,
$$
respectively. Set $\mathcal H=\mathcal I_X\mathcal I_Y$. By Lemma \ref{Hermite_chawla}, $\mathcal Hf$ coincides with the local tensor-product Hermite interpolant and incorporates the mixed derivative data.
\begin{lem}\cite{agarwal1994explicit}
 Suppose that $\partial_x^{R_x+1}f,\partial_y^{R_y+1}f\in L^2(\mathbb R^2)$, then there exist constants $C_X,C_Y>0$, independent of $\nu$ and $\mu$, such that
\begin{align}\label{Hermite_local1}
    \|(I-\mathcal I_X)f\| \leq C_X(\delta_X^m)^{R_x+1} \|\partial_x^{R_x+1}f\|,
\end{align}
and
\begin{align}\label{Hermite_local2}
     \|(I-\mathcal I_Y)f\| \leq C_Y(\delta_Y^n)^{R_y+1} \|\partial_y^{R_y+1}f\|.
\end{align}
The constants $C_X$ and $C_Y$ depend only on $m,n,r,s$ and on the normalized geometry of the sampling sets.
\end{lem}

\begin{lem}\cite{agarwal1994explicit}\label{Hermite_global}
Let $\mathcal H = \mathcal I_X\mathcal I_Y.$
Then, 
\begin{align}\label{Local_hermite_thm}
\|f-\mathcal Hf\|\le{}&C_X(\delta_X^m)^{R_x+1}\|\partial_x^{R_x+1}f\|\nonumber\\
&+C_Y(\delta_Y^n)^{R_y+1}\|\partial_y^{R_y+1}f\|\nonumber\\
&+C_XC_Y(\delta_X^m)^{R_x+1}(\delta_Y^n)^{R_y+1}
\|\partial_x^{R_x+1}\partial_y^{R_y+1}f\|.
\end{align}
for every $f$ for which the derivatives on the right-hand side are in $L^2(\mathbb R^2)$.
\end{lem}

The estimate in Lemma \ref{Hermite_global} is expressed in terms of higher-order derivatives of $f$. To obtain a reconstruction condition directly in terms of $\|f\|$, we next derive a Bernstein-type inequality for functions in $V_2(\phi)$.
\subsection{Bernstein-Type Inequality}
Let $\bm \omega=(\omega_1,\omega_2)$. Now define
\begin{align*}
B_\phi^{(k,l)}(\bm\omega)=\frac{1}{G_\phi(\bm\omega)}\sum_{\bm{n}=(n_1,n_2)\in\mathbb Z^2}
&|\omega_1+n_1|^{2k}|\omega_2+n_2|^{2l}\\[-0.2em]
&\times|\widehat\phi(\bm\omega+\bm{n})|^2.
\end{align*}
and
$$ \Theta_\phi^{(k,l)} = \esssup_{\bm\omega\in[0,1]^2} B_\phi^{(k,l)}(\bm\omega).$$
\begin{lem}[\bf Bernstein-type inequality]\label{berns}
Let $\phi$ be a stable generator of $V_2(\phi)$ such that the derivative of order $(k,l)$ exists in $L^2$ and $\Theta_\phi^{(k,l)}<\infty$. Then every $f\in V_2(\phi)$ satisfies
\begin{eqnarray}\label{bern}
\|\mathcal D^{(k,l)}f\|\leq (2\pi)^{k+l}\, \sqrt{\Theta^{(k,l)}_{\phi}}\,\|f\|,
\end{eqnarray}
for every $f\in V_2(\phi)$. 
\end{lem}
\begin{proof}
Write $f(\mathbf t)=\sum_{\bm{n}}c[\bm{n}]\phi(\mathbf t-\bm{n})$ and
$C(\bm\omega)=\sum_{\bm{n}}c[\bm{n}]e^{-2\pi\mathrm j\bm{n}\cdot\bm\omega}$.
By Plancherel's theorem and periodization over $[0,1]^2$,
\begin{align*}
&\|\mathcal D^{(k,l)}f\|^2\\
&=(2\pi)^{2(k+l)}\int_{[0,1]^2}|C(\bm\omega)|^2\\[-0.2em]
&\,\times\sum_{\bm{n}=(n_1,n_2)\in\mathbb Z^2}|\omega_1+n_1|^{2k}|\omega_2+n_2|^{2l}
|\widehat\phi(\bm\omega+\bm{n})|^2\,d\bm\omega\\
&\le (2\pi)^{2(k+l)}\Theta_\phi^{(k,l)}
\int_{[0,1]^2}|C(\bm\omega)|^2\, G_\phi(\bm\omega)\,d\bm\omega\\
&=(2\pi)^{2(k+l)}\Theta_\phi^{(k,l)}\|f\|^2.
\end{align*}
Taking square roots gives \eqref{bern}.
\end{proof}

\subsection{Approximation operator}
The Hermite interpolant $\mathcal Hf$ is constructed from the measurements, but it does not generally belong to $V_2(\phi)$. We therefore project it back onto the signal space. Let
$$\mathcal P:L^{2}\left({\mathbb{R}}^{2}\right)\longrightarrow V_2\left(\phi \right)$$
be the orthogonal projection onto $V_2\left(\phi \right)$. For a stable generator, the projection can be expressed in the Fourier domain as
\begin{align}\label{Projection_opt}
    \widehat{\mathcal Pg}\left(\bm\omega \right)=\widehat{\phi }\left(\bm \omega \right)\frac{\sum\limits_{\bm{n}\in {\mathbb{Z}}^{2}} \widehat{g}\left(\bm{\omega }+\bm{n}\right)\overline{\widehat{\phi }\left(\bm{\omega} +\bm{n}\right)}}{\sum\limits_{\bm{n}\in {\mathbb{Z}}^{2}} \mid \widehat{\phi }\left(\bm{\omega}+\bm{n}\right){\mid }^{2}},~ g\in L^2(\mathbb{R}^2).
\end{align}
Let
$\mathcal H=\mathcal I_X \mathcal I_Y $
be the global piecewise Hermite interpolation operator introduced in Lemma \ref{Hermite_global}, and define the approximation operator
\begin{align}\label{Approximation_opt}
    \mathcal Af=\mathcal P\mathcal{H}f.
\end{align}
Since $\mathcal Pf=f$ for every $f\in V_2\left(\phi \right)$ and $\mathcal P$ is an orthogonal projection,
$$\| f-\mathcal Af\|=\|\mathcal P\left(f-\mathcal{H}f\right)\|\leq \| f-\mathcal{H}f\|.$$
Combining Lemmas \ref{Hermite_global} and \ref{berns}, define
\begin{align}
    q_{X}&=C_{X}{\left(2\pi \right)}^{R_{x}+1}{\left({\delta }_{X}^{m}\right)}^{R_{x}+1}\sqrt{{\Theta }_{\phi }^{\left(R_{x}+1,\;0\right)}},\\
q_{Y}&=C_{Y}{\left(2\pi \right)}^{R_{y}+1}{\left({\delta }_{Y}^{n}\right)}^{R_{y}+1}\sqrt{{\Theta }_{\phi }^{\left(0,\; R_{y}+1\right)}},
\end{align}
and
\begin{align*}
q_{XY}={}&C_XC_Y(2\pi)^{R_x+R_y+2}(\delta_X^m)^{R_x+1}(\delta_Y^n)^{R_y+1}\\[-0.2em]
&\times\sqrt{\Theta_\phi^{(R_x+1,R_y+1)}}.
\end{align*}
Let $q=q_{X}+q_{Y}+q_{XY}$. Then
$$\| I-A{\| }_{V_2\left(\phi \right)\rightarrow V_2\left(\phi \right)}\leq q.$$

Thus, if $q<1$, the approximation error is contractive. This immediately gives the main iterative reconstruction result.

\begin{thm} [\bf Iterative reconstruction from irregular derivative samples]\label{Iterative_algo_thm}
Let $X={\left\{{\xi }_{i}\right\}}_{i\in \mathbb{Z}}$ and $Y={\left\{{\eta }_{j}\right\}}_{j\in \mathbb{Z}}$ be separated sets with finite block gaps ${\delta }_{X}^{m}$ and ${\delta }_{Y}^{n}$. Let $\phi$ be a stable generator of $V_2\left(\phi \right)$, assume that the required derivative evaluations are bounded on $V_2(\phi)$,  and
$${\Theta }_{\phi }^{\left(R_{x}+1,\;0\right)},\,\ {\Theta }_{\phi }^{\left(0,\; R_{y}+1\right)},\,\ {\Theta }_{\phi }^{\left(R_{x} +1,\;R_{y}+1\right)}$$
are finite. If 
\begin{align}\label{Iterative_algo_condition}
    q=q_{X}+q_{Y}+q_{XY}< 1,
\end{align}
then every $f\in V_2\left(\phi \right)$ can be reconstructed from the irregular derivative samples
$$\Lambda =\left\{\mathcal{D}^{\left(k,\; l\right)}f\left({\xi }_{i},\; {\eta }_{j}\right),\; 0\leq k\leq r,\, 0\leq l\leq s,~i, j\in \mathbb{Z}\right\}$$
through the iterative scheme
\begin{align*}
    f_{0}&=Af,\\
   f_{\nu +1}&=f_{\nu }+A\left(f-f_{\nu }\right), \nu \geq 0. 
\end{align*}
Furthermore,
$\lim\limits_{\nu \rightarrow \infty }f_{\nu }=f$
in $L^{2}({\mathbb{R}}^{2})$, and
$$\| f-f_{\nu }\|\leq q^{\nu +1}\| f\|.$$
\end{thm}
\begin{IEEEproof}
Let $e_{\nu }=f-f_{\nu }.$
Since $f_{0}=Af$,
$e_{0}=\left(I-A\right)f.$
Furthermore,
\begin{align*}
e_{\nu +1}&=f-f_{\nu }-A\left(f-f_{\nu }\right)=\left(I-A\right)e_{\nu }.
\end{align*}
Hence
$$e_{\nu }={\left(I-A\right)}^{\nu +1}f.$$
Since $\|I-A\|\le q<1$, we have
$\| e_{\nu }\|\leq q^{\nu +1}\| f\|.$
Therefore $f_{\nu }\rightarrow f$ in $L^{2}\left({\mathbb{R}}^{2}\right)$. 
\end{IEEEproof}

\begin{rem}\label{last_rem}
If $$f_{\nu +1}=f_{\nu }+\tau A\left(f-f_{\nu }\right),\quad 0<\tau\le 1,$$
then
$$\|I-\tau A\|\le 1-\tau+\tau q=1-\tau (1-q)<1.$$
\end{rem}

\subsection{Numerical Experiments}
The following experiments illustrate the convergence behavior of the proposed iterative reconstruction algorithm. The theoretical condition \eqref{Iterative_algo_condition} is sufficient; the experiments are intended to demonstrate numerical performance and are not used to verify the sharpness of the theoretical density bound.

In the following simulations, the constants $C_X$ and $C_Y$ are evaluated from the normalized one-dimensional Hermite interpolation error operators corresponding to the sampling geometry.
\subsubsection*{\bf Simulation 1} 
\begin{figure}[!t]
    \centering
\includegraphics[width=\columnwidth]{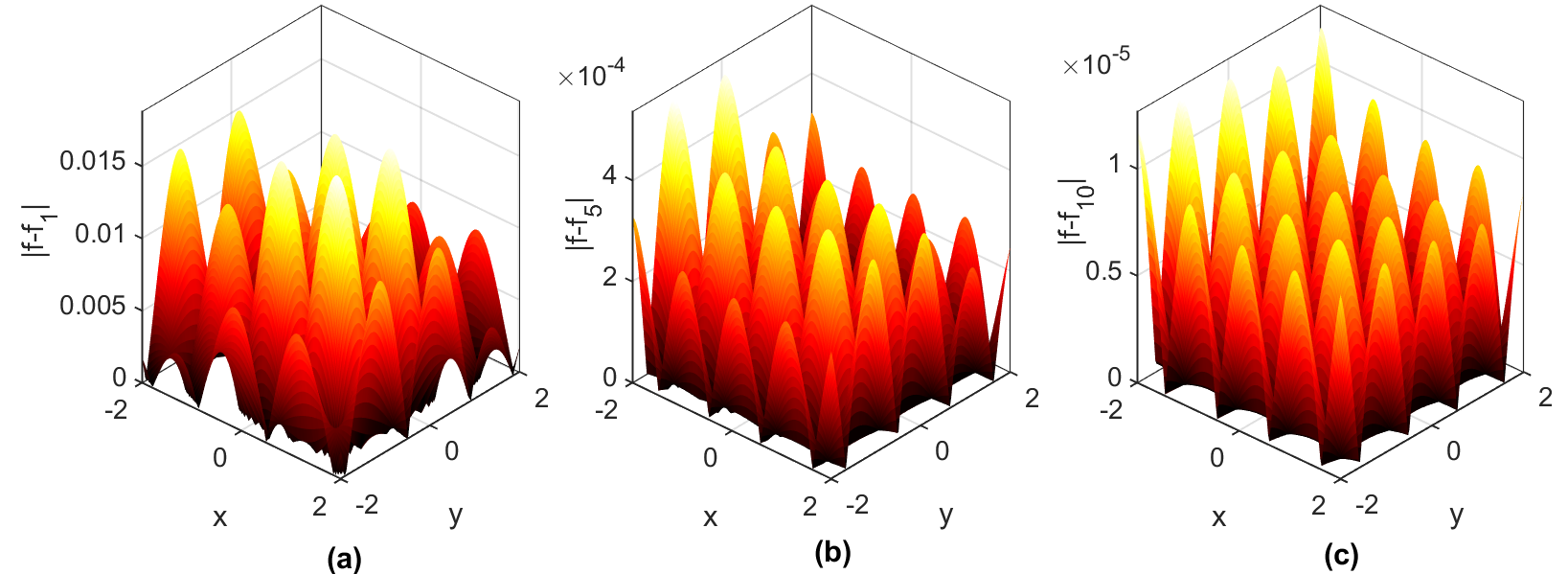}
\caption{Graphical illustration of the reconstruction error $|f-f_\nu|$ for $\nu=1,5,10$. (a) For $\nu=1$, the local relative error is $0.01963$, (b) $\nu=5$, the local relative error is $0.0005312$, and (c) $\nu=10$, the local relative error is $0.0000136$.}
    \label{Simulation1}
\end{figure}

We first consider a function 
$$f(x,y)=\sinc(x)\sinc(y)+\sinc(x-1)\sinc(y+1)\in V_2(\phi),$$
where $\phi(x,y)=\sinc(x)\sinc(y)$. Further assume the sample sets $X=0.15+0.35\mathbb{Z}$, and $Y=0.1+0.4\mathbb{Z}$. Here, $r=s=1$, $m=n=2$, with $\delta_X^2=0.70$ and $\delta_Y^2=0.80$. For the present sampling geometry, the normalized Hermite error constants are $C_X=C_Y\approx2.8404\times10^{-6}$. Using the Bernstein quantities associated with the sinc generator, we obtain $q_X\approx3.21\times10^{-4}$, $q_Y\approx7.16\times10^{-4}$, and $q_{XY}\approx2.30\times10^{-7}$. Consequently,
$q\approx 1.04\times 10^{-3}<1$, We now reconstruct $f$ from the sample set
$$\Lambda=\{\mathcal D^{(k,l)}f(\xi,\eta): ~\xi\in X, \, \eta\in Y, \, 0\le k\le 1,\, 0\le l\le 1\}$$
using the iteration algorithm provided in Theorem \ref{Iterative_algo_thm}. The reconstructed surfaces are displayed over the regions $[-2,2]^2$. Figure \ref{Simulation1} shows the decay of the pointwise error $|f-f_\nu|$ for $\nu=1,5,10$ with $\tau=1$ in Remark \ref{last_rem}, and we also find the local relative error on $[-2,2]^2$.

\subsubsection*{\bf Simulation 2} 
\begin{figure}[!htbp]
    \centering
\includegraphics[width=\columnwidth]{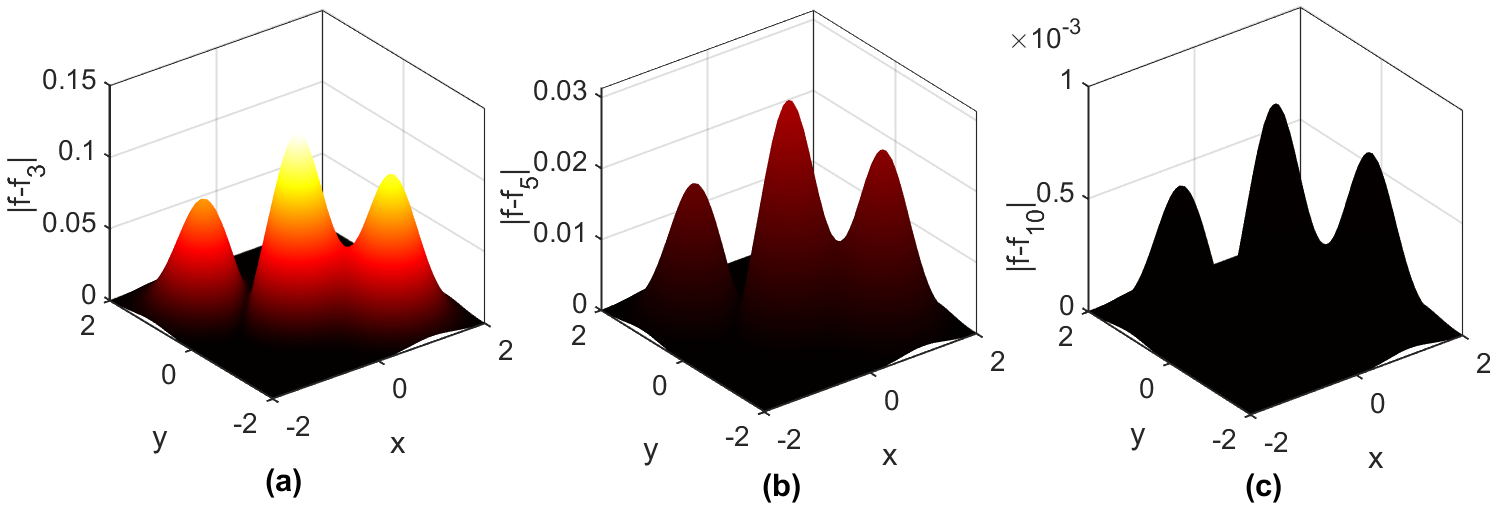}
    \caption{Graphical illustration of the reconstruction error $|f-f_\nu|$ for $\nu=3,5,10$. (a) $\nu=3$, the local relative error is $0.12506$, (b) $\nu=5$, the local relative error is $0.031274$, and (c) $\nu=10$, the local relative error is $0.00097806$.}
    \label{Simulation2}
\end{figure}

As a second example, we consider the stable generator $\phi(x,y)=e^{-\pi(x^2+y^2)}$ and a function 
\begin{multline*}
f(x,y)=\phi(x,y)+0.8\phi(x-1,y+1)\\
-0.6\phi(x+1,y-1)\in V_2(\phi).
\end{multline*}
Further assume the sample sets $X=0.27\mathbb{Z}$, and $Y=0.25+0.59\mathbb{Z}$. Here, $r=1$, $s=2$, $m=1$, $n=2$, with $\delta_X^1=0.27$ and $\delta_Y^2=1.18$. For this configuration, the normalized Hermite error constants are $C_X\approx2.0163\times10^{-3}$ and $C_Y\approx2.6674\times10^{-10}$. Numerical evaluation of the corresponding Bernstein quantities gives $q_X\approx1.15\times10^{-3}$, $q_Y\approx1.30\times10^{-3}$, and $q_{XY}\approx1.49\times10^{-6}$. Hence
$q\approx2.44\times10^{-3}<1,$
so the sufficient convergence condition \ref{Iterative_algo_condition} is satisfied. We now reconstruct $f$ from the sample set
$$\Lambda=\{\mathcal D^{(k,l)}f(\xi,\eta): ~\xi\in X, \, \eta\in Y,\, 0\le k\le 1,\, 0\le l\le 2\}$$
using the iteration algorithm provided in Theorem \ref{Iterative_algo_thm}. The pointwise reconstruction errors for $\nu=3,5,10$ with $\tau=0.5$ in Remark \ref{last_rem} are displayed over $[-2,2]^2$ in Figure \ref{Simulation2}.

\subsection{Robustness to measurement noise and sampling jitter}

The preceding experiments consider exact derivative measurements. We next investigate the practical behavior of the iterative method in the presence of additive measurement noise and perturbations of the sampling locations. These experiments are intended as a numerical robustness study and do not introduce additional theoretical stability assumptions.

\begin{figure}[!t]
    \centering
\includegraphics[width=\columnwidth]{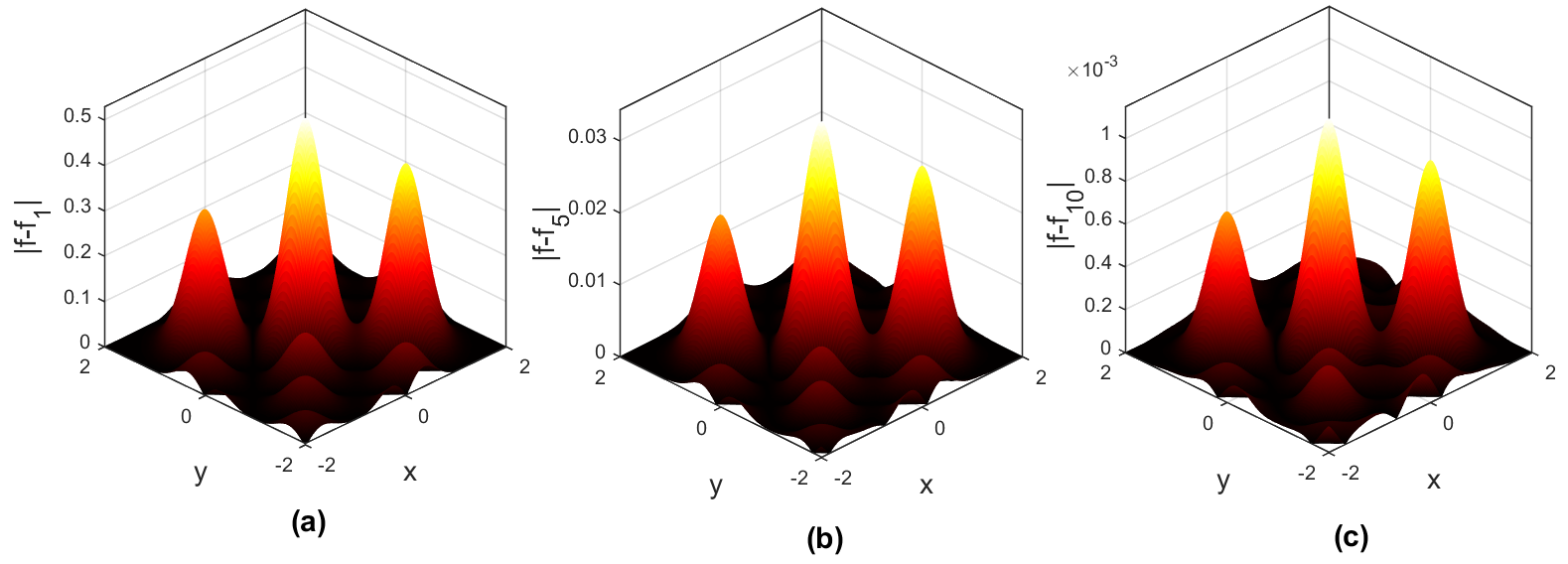}
    \caption{Graphical illustration of the reconstruction error $|f-f_\nu|$, for $\nu=1, 5, 10$. (a) $\nu=1$, the local relative error is $0.50342$, (b) $\nu=5$, the local relative error is $0.032543$, and (c) $\nu=10$, the local relative error is $0.0010825$.}
    \label{fig:SNR_image1}
\end{figure}

We consider the stable generator
$\phi(x,y)=\operatorname{sinc}(x+y)e^{-\pi(x-y)^2}$ and the function
\begin{multline*}
f(x,y)=\phi(x,y)+0.8\phi(x-1,y+1)\\
-0.6\phi(x+1,y-1)\in V_2(\phi).
\end{multline*}
with $X=0.27\mathbb{Z}$, and $Y=0.25+0.59\mathbb{Z}$. We now reconstruct $f$ from the sample set
\begin{multline*}
\Lambda=\{\mathcal D^{(k,l)}f(\xi_i,\eta_j):\xi_i\in X,\eta_j\in Y,\\
0\le k\le1,\ 0\le l\le2\}.
\end{multline*} 
We first examine the reconstruction error in the noise-free setting. Figure \ref{fig:SNR_image1} displays the pointwise error $|f-f_\nu|$ at selected iterations.

\noindent
\textbf{Additive measurement noise.} To examine the robustness of the iterative reconstruction method with respect to measurement noise, we perturb the exact derivative samples according to
$$ \widetilde{y}_{i,j}^{(k,l)} = \mathcal{D}^{(k,l)}f(\xi_i,\eta_j) + \varepsilon_{i,j}^{(k,l)}, $$
where $\varepsilon_{i,j}^{(k,l)}$ denotes zero-mean Gaussian noise. Let $\mathbf y$ denote the vector obtained by stacking all exact derivative measurements and let $\boldsymbol{\varepsilon}$ denote the corresponding noise vector. The measurement signal-to-noise ratio is defined by
$$ \mathrm{SNR} = 20\log_{10} \frac{\|\mathbf y\|} {\|\boldsymbol{\varepsilon}\|}. $$

We consider the noise levels
$ \mathrm{SNR}=20,\;30,\;40\ {\rm dB}. $
For each SNR level, the experiment is repeated over $50$ independent Gaussian noise realizations, and the local relative $L^2$-reconstruction errors are averaged over the trials.

\begin{figure}[!t]
    \centering
\includegraphics[width=3in]{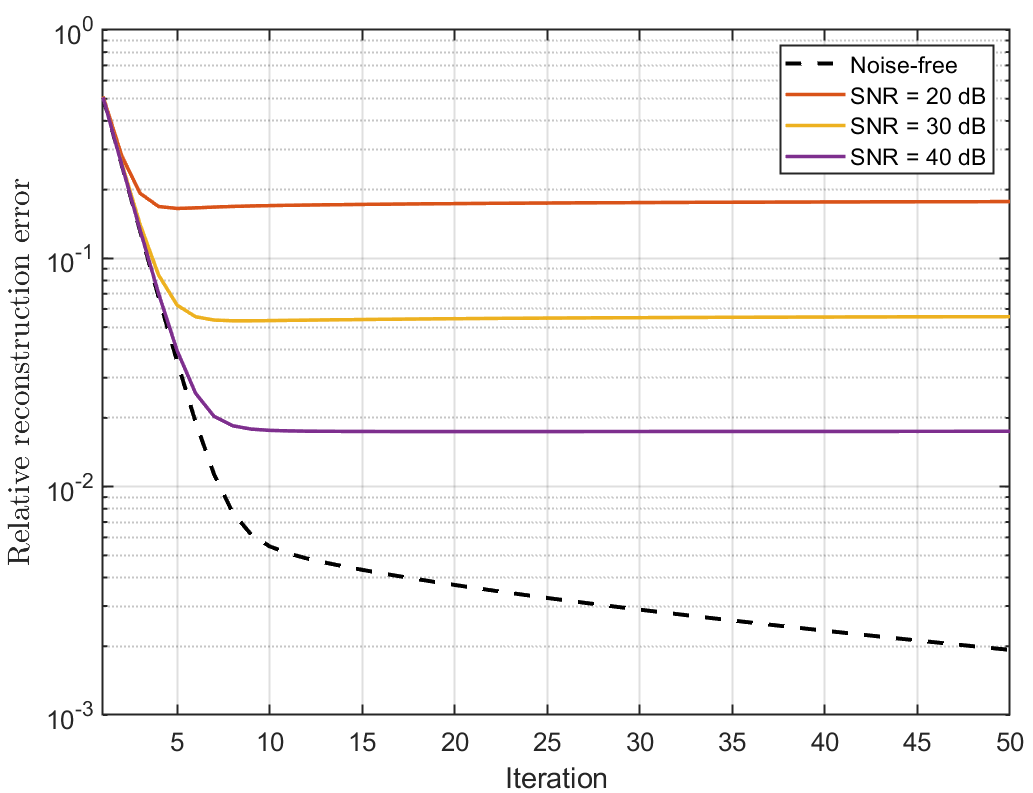}
    \caption{Relative $L^2$-reconstruction error versus iteration for measurement
SNRs of 20, 30, and 40 dB. The curves are averaged over 50 independent
Gaussian noise realizations.}
    \label{fig:SNR_image2}
\end{figure}

Figure \ref{fig:SNR_image2} shows rapid initial decay followed by an SNR-dependent error floor. We also report the normalized mean-square error
$$ \mathrm{NMSE}_{\mathrm{dB}} = 10\log_{10}\left(\frac{\|f-f_\nu\|^2}{\|f\|^2}\right). $$

At $\nu=25$, NMSE is averaged over $50$ independent trials on the linear scale and then converted to decibels.
\begin{table}[!t]
\centering
\caption{NMSE averaged before conversion to decibels for $\nu=25$.}
\label{tab:snr_nmse}
\begin{tabular}{c c c c}
\hline
\textbf{SNR (dB)} & \textbf{Mean NMSE} & \textbf{NMSE (dB)} & \textbf{Std(NMSE)} \\
\hline
10 & 0.094535   & -10.244 & 0.028664 \\
20 & 0.0091914  & -20.366 & 0.0028824 \\
30 & 0.0009596  & -30.179 & 0.00027672 \\
40 & $9.4569\times 10^{-5}$ & -40.243 & $2.8892\times 10^{-5}$ \\
\hline
\end{tabular}
\end{table}

\textbf{Sampling jitter.} We next investigate perturbations in the sampling locations. Instead of the ideal sampling points $(\xi_i,\eta_j)$, we use perturbed locations
$$\widetilde{\xi}_i=\xi_i+\varepsilon_i^x, \quad \widetilde{\eta}_j=\eta_j+\varepsilon_i^y,$$
where the perturbations $\varepsilon_i^x$ and $\varepsilon_i^y$ are generated independently and satisfy
$|\varepsilon_i^x| \leq \rho \Delta \xi$ and $|\varepsilon_i^y| \leq \rho \Delta\eta.$
Here, $\Delta\xi$ and $\Delta\eta$ denote the sampling spacings in the $x$- and $y$-directions, respectively, and
$\rho \in \{0.01, 0.02, 0.05, 0.10\}.$
We consider $1\%,2\%,5\%$, and $10\%$ perturbations of the nominal intervals. For each level, derivative samples are evaluated at the perturbed locations and reconstruction uses the nominal geometry. 
\begin{table}[!t]
\centering
\caption{Local relative error: mean $\pm$ sample standard deviation.}
\begin{tabular}{c c c c}
\hline
\textbf{Jitter} 
& \textbf{Iter 10} 
& \textbf{Iter 25} 
& \textbf{Iter 50} \\
\hline
1\%  & $0.0045 \pm 0.0016$ & $0.0044 \pm 0.0015$ & $0.0044 \pm 0.0015$ \\
2\%  & $0.0094 \pm 0.0029$ & $0.0092 \pm 0.0028$ & $0.0092 \pm 0.0028$ \\
5\%  & $0.0216 \pm 0.0068$ & $0.0215 \pm 0.0067$ & $0.0215 \pm 0.0067$ \\
10\% & $0.0404 \pm 0.0131$ & $0.0403 \pm 0.0131$ & $0.0403 \pm 0.0131$ \\
\hline
\end{tabular}
\label{tab:jitter_results}
\end{table}
Table \ref{tab:jitter_results} summarizes the corresponding reconstruction errors.

\subsection{Reconstruction of an image-generated signal with a non-separable generator}

\begin{figure}[!t]
    \centering
\includegraphics[width=\columnwidth]{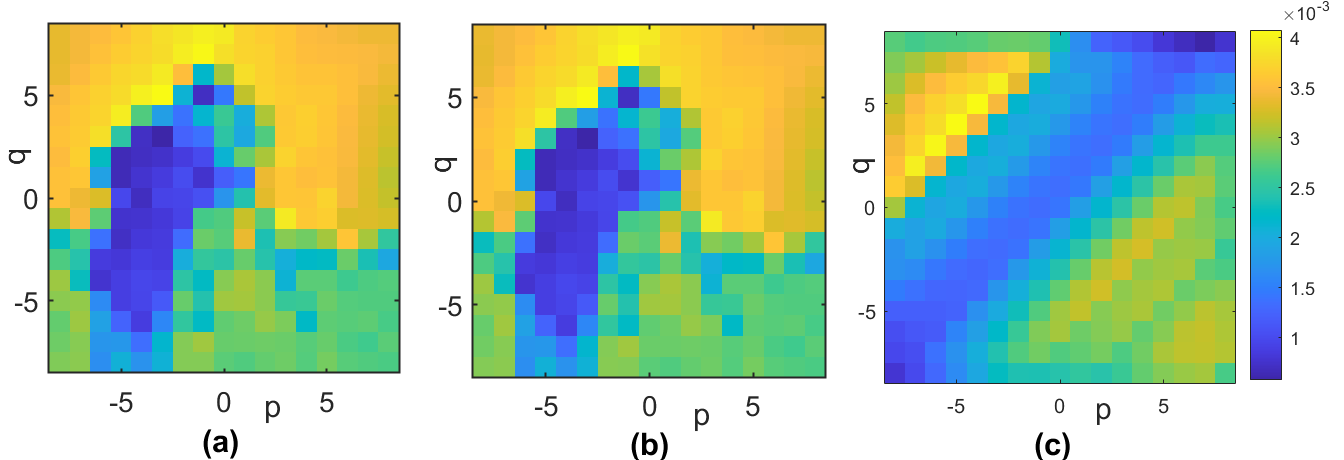}
    \caption{(a) The target lattice samples $f(p,q)$, (b) the reconstructed samples $f_{\mathrm{rec}}(p,q)$ after $50$ iterations, and (c) the corresponding absolute error.}
    \label{fig:Image_Processing1}
\end{figure}

To illustrate the applicability of the proposed iterative method to two-dimensional signals, we consider the non-separable generator
$$ \phi(x,y)=\operatorname{sinc}(x+y)e^{-\pi(x-y)^2}. $$
The standard Cameraman image is resized to a $17\times17$ grayscale array, and its $289$ pixel intensities are used as the coefficient sequence $c[p,q]$, $p,q=-8,\ldots,8$. We define the continuous target signal
$$ f(x,y) = \sum_{p=-8}^{8}\sum_{q=-8}^{8} c[p,q]\phi(x-p,y-q), $$
so that $f\in V_2(\phi)$ by construction. We use the derivative measurements $f$, $f_y$, $f_{yy}$, $f_x$,  $f_{xy}$, $f_{xyy}$
on the sampling sets
$X=0.27\mathbb Z$ and $Y=0.25+0.59\mathbb Z.$
At each iteration, the residual derivative measurements are locally interpolated using the bivariate Hermite operator, and the resulting correction is projected onto $V_2(\phi)$. Figure \ref{fig:Image_Processing1} shows the target samples, the reconstruction after 50 iterations, and the corresponding pointwise absolute error.

For image-quality assessment, we additionally use the peak signal-to-noise ratio (PSNR) and structural similarity index (SSIM). If $I$ denotes the reference image and $\widetilde I$ its reconstruction, then
$$ \operatorname{MSE} = \frac1N\sum_{j=1}^{N}|I_j-\widetilde I_j|^2, \quad \operatorname{PSNR} = 10\log_{10} \left( \frac{L^2}{\operatorname{MSE}} \right), $$
where $L$ is the reference-image dynamic range. SSIM \cite{wang2004image} and PSNR are evaluated between $\{f(p,q)\}$ and $\{f_{\mathrm{rec}}(p,q)\}$.

\begin{figure}[!htbp]
    \centering
\includegraphics[width=2in]{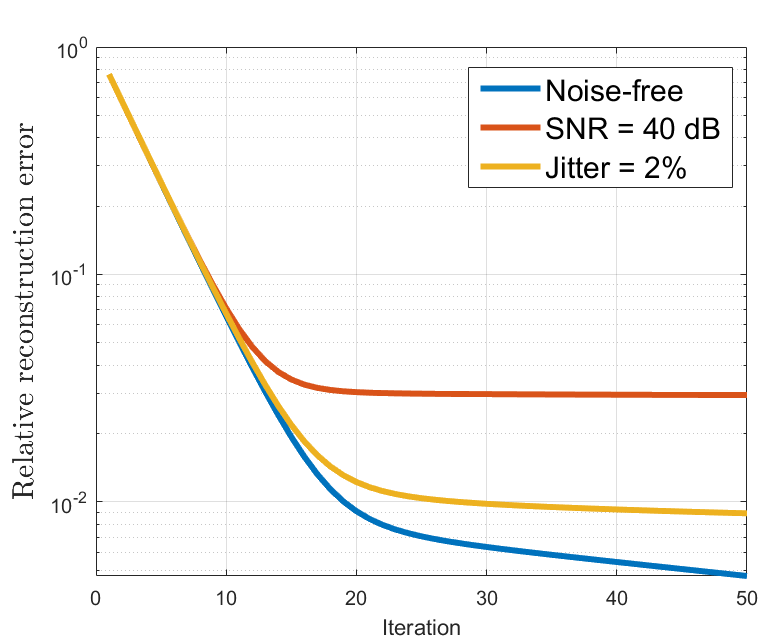}
    \caption{Relative $L^2$-reconstruction error versus iteration for noise-free measurements, 40 dB additive measurement noise, and $2\%$ sampling jitter. The noisy and jittered curves are averaged over 50 independent realizations.}
    \label{fig:Image_Processing2}
\end{figure}
Figure $\ref{fig:Image_Processing2}$ shows the convergence behavior under exact measurements, $40$ dB additive measurement noise, and a $2\%$ perturbation of the sampling locations. The corresponding reconstruction performance is summarized in Table $\ref{tab:reconstruction_performance}$:
\begin{table}[!t]
\centering
\caption{Image-reconstruction performance under different conditions. The relative $L^2$-error is computed over the evaluation region.}
\setlength{\tabcolsep}{3pt}
\begin{tabular}{lccc}
\hline
 & \textbf{Relative error} & \textbf{PSNR (dB)} & \textbf{SSIM} \\
\hline
Noise-free
& $0.00473$
& $52.693$ 
& $0.99998$ \\

40 dB measurement noise
& $0.02951$
& $34.536$
& $0.96818$ \\

$2\%$ sampling jitter
& $0.00891$
& $44.836$ 
& $0.99791$ \\
\hline
\end{tabular}
\label{tab:reconstruction_performance}
\end{table}

\FloatBarrier
\section{Conclusion}
We developed a multidimensional multichannel sampling framework for principal SISs based on a finite-dimensional polyphase representation. Periodic nonuniform, derivative, and filtered measurements are handled in a common formulation, with stable recovery characterized by uniform singular-value bounds and synthesis kernels obtained from the Moore--Penrose left inverse. The approach applies to separable and genuinely non-separable generators. For irregular derivative sampling, we proposed a local Hermite-interpolation/projection iteration and derived a sufficient contraction condition for geometric convergence. Numerical experiments demonstrate the behavior of both approaches under exact measurements, additive noise, and sampling-location perturbations.

\bibliographystyle{IEEEtran}
\bibliography{References}
\end{document}